\documentclass[11pt,dvips]{article}

\usepackage{epsfig}
\usepackage{graphicx,float}
\usepackage{array}
\usepackage{fancybox}
\usepackage{tabularx}
\usepackage{indentfirst}
\usepackage{amsmath,amssymb,bbold}
\usepackage{url}
\usepackage{rotating}
\usepackage{color}
\usepackage{hhline}
\usepackage{hyperref}
\usepackage{slashed}
\usepackage{tikz}
\usepackage{makecell}
\usepackage{pgfplots}
\usepackage{empheq}
\usepackage{physics}
\usepackage{float}

\restylefloat{table}

\newcommand{\s}{\\ \vspace*{-3.5mm}}
 \newcommand{\rb}[2]{\raisebox{#1}[-#1]{#2}}
 \newcommand{\sts}{\scriptstyle}

\renewcommand{\thefootnote}{\alph{footnote}}

\begin{document}


\begin{flushright}
\end{flushright}

\vskip 1.8cm

\begin{center}
{\LARGE \bf Vector currents of integer-spin Majorana particles}\\[1.0cm]
{Seong Youl Choi\footnote{sychoi@jbnu.ac.kr} and
 Jae Hoon Jeong\footnote{jaehoonjeong229@gmail.com}} \\[0.5cm]
{\it  Department of Physics and RIPC, Jeonbuk National University,
      Jeonju 54896, Korea}
\end{center}

\vskip 1.5cm

\begin{abstract}
\noindent
A general and comprehensive analysis for the vector currents of two
massive particles, $X_2$ and $X_1$, with arbitrary integer-spin values
is given. Our special focus is on the case when two particles are
charge self-conjugate, i.e. Majorana bosons. The general structure
of their couplings to an on-shell or off-shell vector boson $V$ is
described in a manifestly covariant way and then the constraints on
the triple vertex due to discrete CP symmetry and the Majorana condition
of two particles being Majorana are worked out. The validity of our full
analytic investigation is checked by studying the two-body decay,
$X_2\to V X_1$, with an on-shell or off-shell $V$ boson in the helicity
formalism complementary to the covariant formulation. Threshold effects
of the two-lepton invariant-mass and polar-angle correlations in the two
sequential two-body decays, $X_2\to V X_1$ and $V\to \ell^-\ell^+$ with
$\ell=e$ or $\mu$, are derived analytically in a compact form by use
of the Wick helicity rotation and they are investigated numerically
in a few specific spin-combination scenarios for probing the spin and
dynamical structure of the $X_2X_1V$ vertex.
\end{abstract}



\vskip 1.5cm

\renewcommand{\thefootnote}{\fnsymbol{footnote}}

\section{Introduction}
\label{sec:introduction}

The Standard Model (SM)~\cite{Glashow:1961tr,Weinberg:1967tq,Salam:1968rm}
and beyond contain several particles that are identical to their own antiparticles.
In the following, for the sake of a unified description those charge
self-conjugate particles are called Majorana bosons or fermions, depending on
whether their spins are integer or half-integer, although the term Majorana
was used for a charge self-conjugate spin-1/2 fermion originally introduced
by Majorana~\cite{Majorana:1937vz} through the formulation of a purely real
version of the Dirac equation~\cite{Dirac:1928hu}. \s

Representatively, the spin-0 elementary SM Higgs boson $H$ discovered at the
CERN Large Hadron Collider in 2012~\cite{Aad:2012tfa,Chatrchyan:2012ufa},
various composite mesons such as the spin-0 $\pi^0$, the spin-1 $\rho^0$ and
the spin-2 $a^0_2$ as well as the spin-1 $J/\psi$, the spin-1 SM isospin-neutral
gauge bosons, $\gamma$ and $Z$, and the color-neutral
gluons~\footnote{Conceptually, the term ``color-neutral" is not identical to
the term ``colorless" but it rather means the neutral component of a color
octet under strong gauge symmetry. Similarly, an isospin-neutral $Z$ is
not simply an isospin-singlet but it is a component of a triplet under
electroweak gauge symmetry.}, are Majorana bosons.
The presence of Majorana particles is predicted also by various versions
of grand-unified theories~\cite{Langacker:1980js,Croon:2019kpe} and it is
guaranteed in $N=1$ supersymmetric gauge theories linking bosons to fermions
and vice versa~\cite{Fayet:1976cr,Nilles:1983ge,Haber:1984rc}.
The supersymmetric partners of neutral gauge bosons and Higgs bosons are
Majorana fermions. In addition,  one of the leading unanswered questions
in the context of neutrino physics~\cite{GonzalezGarcia:2007ib}
is whether massive neutrinos are Majorana or Dirac particles. {\it The concept
of Majorana particles is ubiquitous} in nuclear and particle physics and even
in condensed-matter physics~\cite{Elliott:2014iha,Wilczek:2009np5614,
Leggett:2016oci}.\s

Previously, the electromagnetic properties of two identical-spin
particles of possibly different masses and of spin values up to 3/2 have
been investigated extensively~\cite{Schechter:1981hw,Li:1981um,Pal:1981rm,
Halprin:1983ez,Nieves:1982bq,Khare:1983tm,Bilenky:1984fg,Rosen:1983wu,
Kayser:1982br,Kayser:1984ge,Nieves:1996ff,Nieves:2013csa} and those of
two identical Majorana particles of arbitrary spin have been worked out
in detail in Refs.$\,$\cite{Radescu:1985wf,Boudjema:1988zs,Boudjema:1990st}.
In this work, as a natural extension of those previous works and a powerful
platform for probing Majorana particles systematically, {\it we provide
a general analysis of the vector-current interactions of an on-shell
or off-shell vector boson with two massive on-shell integer-spin
particles, of which the masses and spins do not have to be identical, and
then we elaborate on the special properties of the triple vertex when
two particles are Majorana bosons.}\footnote{The general structure of
the currents of three off-shell vector particles was presented and
discussed in Ref.$\,$\cite{Nieves:1996ff}.} \s

The general structure of the vector-current vertex of two Majorana particles
and a vector boson can be probed in various production and decay processes
at hadron colliders and $e^-e^+$ colliders~\cite{Ellis:1983er,Petcov:1984nf,
Bilenky:1985wu,Bilenky:1986nd,Petcov:1986sc,MoortgatPick:2002iq,Khristova:1987xq,
Balantekin:2018ukw,Renard:1981es,Hagiwara:1986vm,Choudhury:1994nt,
Ananthanarayan:2004eb,Rahaman:2016pqj,Rahaman:2017qql,Rahaman:2020jll,
Rahaman:2018ujg,Choi:2015zka,Choi:2001ww,Choi:1994nv,Choi:2003fs,Choi:2005gt,
Choi:2018sqc,Choi:2019aig,Choi:2020spm}. Especially, production of a pair of Majorana particles of arbitrary spins at $e^-e^+$ colliders and/or two-body
(or three-body) decays involving two non-degenerate Majorana particles can serve
as a powerful handle for probing the vertex structure~\cite{Choi:2020spm}.
In the present work, as a straightforward and simple check of the validity
of our general analytic analysis on the vertex, we make a detailed
study of the two-body decay $X_2\to VX_1$ of a heavier Majorana boson $X_2$
into a lighter Majorana boson $X_1$ and an on-shell or off-shell vector
boson $V$.\s

This general and model-independent study of the vector currents of two
massive (Majorana) particles of arbitrary integer spins coupled to an on-shell
or off-shell vector boson can serve as a powerful guide for investigating
various interactions among the SM particles and for finding and characterizing
new physics beyond the SM (BSM). A similar analysis for half-integer spin
particles will be reported separately.\s

The paper is organized as follows. Firstly, we make a systematic derivation
and provide a general analysis of the $X_2X_1V$ vector-current vertex of
two massive on-shell particles, $X_2$ and $X_1$, of arbitrary integer spins
and an on-shell or off-shell vector boson $V$ in a manifestly covariant way
in Section~\ref{sec:z_coupling_two_majorana_bosons}. This general vertex
form is valid irrespective of whether two particles are charged or neutral.
The covariant formulation enables us to efficiently derive all the
Lorentz-covariant terms of the vertex and to systematically analyze
its key characteristics related to discrete CP symmetry or/and the
Majorana condition that two particles are charge self-conjugate,
i.e. Majorana bosons. Secondly, we check the validity of our covariant
description explicitly by analyzing the two-body decay, $X_2\to V X_1$,
with an on-shell or off-shell vector boson $V$ in the helicity formulation
originally developed in Ref.$\,$\cite{Jacob:1959at} and refined and used
in a slightly different form later, for example, in
Refs.$\,$\cite{Wick:1962zz,Leader:2001gr,Chung:1971ri}
in Section~\ref{sec:decay_helicity_amplitudes}. This helicity formalism is
equivalent and complementary to the covariant formalism in that it facilitates
the enumeration of all the independent terms and the symmetry arguments
very efficiently. Thirdly, in combination with the leptonic $V$ decays
$V\to\ell^-\ell^+$ with $\ell=e$ or $\mu$, we investigate the possibility
of determining the spin and dynamical structure of the triple vertex through
(polar-)angle correlations and/or lepton invariant-mass distributions to be
taken into account when the vector boson $V$ has to be unavoidably virtual
in Section~\ref{sec:correlated_angular_distributions}. Finally we summarize
our findings and conclude in Section~\ref{sec:conclusions}.\s

\section{Vertex of two Majorana bosons and a vector boson}
\label{sec:z_coupling_two_majorana_bosons}

An on-shell boson of integer spin $s$, mass $m$, momentum $k$ and
helicity $\lambda$ is defined by a rank-$s$ wave tensor $\epsilon^{\alpha_1\cdots\alpha_s}(k,\lambda)$~\cite{Behrends:1957rup,
Auvil:1966eao,Weinberg:1995mt} that is completely
symmetric, traceless and divergence-free
 \begin{eqnarray}
   \varepsilon_{\mu\nu\alpha_i\alpha_j}\,
   \epsilon^{\alpha_1\cdots\alpha_i\cdots\alpha_j\cdots\alpha_s}(k,\lambda)
&=& 0\,,
   \label{eq:totally_symmetric}\\[1mm]
   g_{\alpha_i\alpha_j}\,
   \epsilon^{\alpha_1\cdots\alpha_i\cdots\alpha_j\cdots\alpha_s}(k,\lambda)
&=& 0\,,
   \label{eq:traceless}\\[1mm]
    k_{\alpha_i}\,
   \epsilon^{\alpha_1\cdots\alpha_i\cdots\alpha_s}(k,\lambda)
&=& 0\,,
   \label{eq:divergence_free}
 \end{eqnarray}
and the wave tensor satisfies the equation $(k^2-m^2)\,
\epsilon^{\alpha_1\cdots\alpha_s}(k,\lambda)=0$ for any helicity value
$\lambda$ taking an integer value between $-s$ and $s$. The wave tensor
can be expressed explicitly by a linear combination of $s$ products of
spin-1 wave vectors with appropriate Clebsch-Gordon coefficients. \s


%
\begin{figure}[H]
\begin{center}
\includegraphics[width=9.0cm, height=5.5cm]{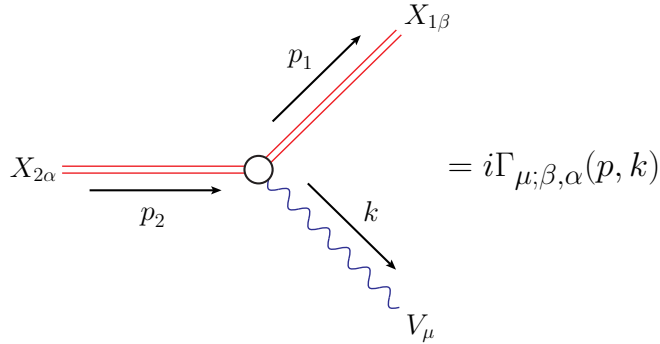}
\caption{\it Feynman rules for the general $X_2X_1V$ vertices of a spin-$s_2$
             particle $X_2$, a spin-$s_1$ particle $X_1$ and a spin-1
             vector boson $V$. The indices, $\beta$ and $\alpha$, stand for
             the sequences of the $s_1$ and $s_2$ indices,
             $\beta=\beta_1\cdots \beta_{s_1}$ and
             $\alpha=\alpha_1\cdots\alpha_{s_2}$,
             collectively. $p=p_2+p_1$ and $k=p_2-p_1$.
}
\label{fig:feynman_rules_x2x1v_vertex}
\end{center}
\end{figure}

The vector currents of two on-shell bosons, $X_2$ of mass $m_2$ and spin
$s_2$ and $X_1$ of mass $m_1$ and spin $s_1$, can be written in a general
form, which is applicable independently of whether the particles are
charged or neutral, as
\begin{eqnarray}
    J^{s_2s_1}_\mu (p,k;\lambda_2,\lambda_1)
&=& \langle X_1(p_1,\lambda_1)|V_\mu| X_2 (p_2,\lambda_2)\rangle
    \nonumber\\
&=& \epsilon_1^{*\beta_1\cdots\beta_{s_1}}(p_1,\lambda_1)\,
  \Gamma_{\mu;\beta_1\cdots\beta_{s_1},\alpha_1\cdots\alpha_{s_2}}(p,k)\,\,
  \epsilon_2^{\alpha_1\cdots\alpha_{s_2}}(p_2,\lambda_2)\,,
\label{eq:vector_vertex_two_majorana_bosons_x2_x1} \\
   \bar{J}^{s_1s_2}_\mu (p,k;\lambda_1,\lambda_2)
&=& \langle X_2(p_2,\lambda_2)|V^\dagger_\mu| X_1 (p_1,\lambda_1)\rangle
    \nonumber\\
&=& \epsilon_2^{*\alpha_1\cdots\alpha_{s_2}}(p_2,\lambda_2)\,
  \bar{\Gamma}_{\mu;\alpha_1\cdots\alpha_{s_2},\beta_1\cdots\beta_{s_1}}(p,k)\,
  \epsilon_1^{\beta_1\cdots\beta_{s_1}}(p_1,\lambda_1)\,,
\label{eq:vector_vertex_two_majorana_bosons_x1_x2}
\end{eqnarray}
for the $X_2\to X_1$ and $X_1\to X_2$ transitions where $p_{1,2}$ and $\lambda_{1,2}$ are the momenta and helicities of the particles, $X_{1,2}$,
respectively. Two independent momenta, $p=p_2+p_1$ and $k=p_2-p_1$, are
introduced for the sake of a systematic and unified description of the
two triple vertices. The Feynman rule of the $X_2X_1V$ interaction vertex
is depicted diagrammatically in Figure~\ref{fig:feynman_rules_x2x1v_vertex}.
If any absorptive parts are ignored, the vector vertex operator $V_\mu$
is Hermitian, i.e. $V^\dagger_\mu=V_\mu$. \s

If the currents $J$ and $\bar{J}$ are coupled to an on-shell vector boson
$V$ such as a photon $\gamma$ and a massive gauge boson $Z$ or to a conserved
vector current through an off-shell $V$ exchange, we can impose
the transversality condition
\begin{eqnarray}
k^\mu\, J^{s_2s_1}_\mu (p,k; \lambda_2,\lambda_1)=0\quad \mbox{and}\quad
k^\mu\, \bar{J}^{s_1s_2}_\mu (p,k; \lambda_1,\lambda_2)=0\,,
\label{eq:transversality_condition}
\end{eqnarray}
with no loss of generality, which effectively kill every term proportional
to $k_\mu$ in the vertices. \s

Utilizing the general properties (\ref{eq:totally_symmetric}),
(\ref{eq:traceless}) and (\ref{eq:divergence_free}) of the wave tensors,
$\epsilon_1(p_1,\lambda_1)$ and $\epsilon_2(p_2,\lambda_2)$,
let us derive the most general form of the $X_2X_1 V$ vertex, $\Gamma_{\mu;\beta_1\cdots\beta_{s_1},\alpha_1\cdots\alpha_{s_2}}(p,k)$,
depicted in Figure~\ref{fig:feynman_rules_x2x1v_vertex}.
[For notational convenience, frequently we use $\beta$ and $\alpha$
collectively standing for the sequences of the indices,
$\beta_1\cdots \beta_{s_1}$ and $\alpha_1\cdots\alpha_{s_2}$.]
In the following, we deal with the identical spin case of $s_2=s_1=s$
and the different spin case of $s_2\neq s_1$ separately. \s

\subsection{Identical spin case: \boldmath{$s_2=s_1=s$}}
\label{subsec:identical_spin_case}

If the $X_2$ and $X_1$ spins are identical, i.e. $s_2=s_1=s$, the most
general form of each of $\Gamma_{\mu;\beta,\alpha}$ and
$\bar{\Gamma}_{\mu;\alpha,\beta}$ can be decomposed in six parts as
\begin{eqnarray}
 \Gamma_{\mu; \beta_1\cdots\beta_{s},\alpha_1\cdots\alpha_{s}}(p,k)
&=& p_\mu\, F^1_{\beta_1\cdots\beta_{s},\alpha_1\cdots\alpha_{s}}(p,k)
  + k_\mu\, F^2_{\beta_1\cdots\beta_{s},\alpha_1\cdots\alpha_{s}}(p,k)
  \nonumber\\
&+& (g_{\mu\beta_1} p_{\alpha_1} +  g_{\mu\alpha_1} p_{\beta_1})\,
     G^1_{\beta_2\cdots\beta_{s},\alpha_2\cdots\alpha_{s}}(p,k)
  \nonumber\\
&+& \langle \mu\beta_1\alpha_1 k\rangle\,
    G^2_{\beta_2\cdots\beta_{s},\alpha_2\cdots\alpha_{s}}(p,k)
  \nonumber\\
&+& (g_{\mu\beta_1}p_{\alpha_1}-g_{\mu\alpha_1}p_{\beta_1})\,
     G^3_{\beta_2\cdots\beta_{s},\alpha_2\cdots\alpha_{s}}(p,k)
  \nonumber\\
&+& \langle \mu\beta_1\alpha_1 p\rangle \,
    G^4_{\beta_2\cdots\beta_{s},\alpha_2\cdots\alpha_{s}}(p,k) \,,
\label{eq:identical_spin_general_decomposition_x2_x1} \\
 \bar{\Gamma}_{\mu; \alpha_1\cdots\alpha_{s},\beta_1\cdots\beta_{s}}(p,k)
&=& \Gamma_{\mu; \beta_1\cdots\beta_{s},\alpha_1\cdots\alpha_{s}}(p,k)
    [\, F^a \to \bar{F}^a,\, G^b \to \bar{G}^b\,]\,,
\label{eq:identical_spin_general_decomposition_x1_x2}
\end{eqnarray}
with the abbreviations, $\langle \mu \beta_1\alpha_1 p\rangle
=\varepsilon_{\mu\beta_1\alpha_1\rho}\, p^\rho$ and
$\langle \mu \beta_1\alpha_1 k\rangle=\varepsilon_{\mu\beta_1\alpha_1\rho}\, k^\rho$,
and the indices, $a=1, 2$ and $b=1,2,3,4$, enumerating all the allowed
terms. [For future reference, we note
here that our convention of the totally antisymmetric Levi-Civita tensor is $\epsilon_{0123}=+1$.]
{\it The totally symmetric wave tensors to be coupled to the vertices in
Eqs.$\,$(\ref{eq:identical_spin_general_decomposition_x2_x1})
and (\ref{eq:identical_spin_general_decomposition_x1_x2})
guarantee the automatic symmetrization of all the terms including the tensors,
$F^a(p,k)$, $G^b(p,k)$, $\bar{F}^a(p,k)$ and $\bar{G}^b(p,k)$, under
any $\alpha$-index and/or $\beta$-index permutations.}~\footnote{If the
vector boson $V$ couples to a conserved vector current, then the terms
with $k_\mu$ in Eqs.$\,$(\ref{eq:identical_spin_general_decomposition_x2_x1})
and (\ref{eq:identical_spin_general_decomposition_x1_x2}) do not
contribute to the $X_2\leftrightarrow X_1$ transitions. The same
argument is valid even in the different spin case to be discussed
in the next subsection.} \s

While the $X_2\to X_1$ and $X_1\to X_2$ transition vertices of two spinless
particles of $s_2=0$ and $s_1=0$ have the contribution only from the
first two parts in each of Eqs.$\,$(\ref{eq:identical_spin_general_decomposition_x2_x1})
and (\ref{eq:identical_spin_general_decomposition_x1_x2}),
the remaining four parts in each equation start participating in constructing
the vertices of two Majorana bosons with spin $s\geq 1$.\s

One crucial point to be exploited for organizing all the independent terms
contributing to the triple vertices is that both
$p_{\beta} \langle \mu\alpha pk\rangle$ and
$p_{\alpha} \langle \mu\beta pk\rangle$ for any 4-vector indices,
$\mu,\,\alpha$ and $\beta$, can be
expressed in terms of other tensor terms so that they are not
independent any more. This can easily be seen as follows. Since
no rank-5 completely antisymmetric tensor exists in four dimensions,
the following identity holds:
\begin{eqnarray}
 g_{\lambda\mu}\varepsilon_{\alpha\beta\rho\sigma}
-g_{\lambda\alpha}\varepsilon_{\mu\beta\rho\sigma}
+g_{\lambda\beta}\varepsilon_{\mu\alpha\rho\sigma}
-g_{\lambda\rho}\varepsilon_{\mu\alpha\beta\sigma}
+g_{\lambda\sigma}\varepsilon_{\mu\alpha\beta\rho} =0\,.
\label{eq:schouten_identity}
\end{eqnarray}
By multiplying the above equation by $k^\lambda p^\rho k^\sigma$ and
$p^\lambda p^\rho k^\sigma$, we find
\begin{eqnarray}
  p_\beta \langle \mu\alpha pk\rangle
 +p_\alpha \langle \mu\beta pk\rangle
&=& k_\mu\, \langle \beta\alpha pk\rangle
 +k^2\, \langle \mu\beta\alpha p \rangle
 -p.k\, \langle \mu\beta\alpha k\rangle
  \,,
\label{eq:redundant_terms_sum} \\
  p_\beta \langle \mu\alpha pk\rangle
 -p_\alpha \langle \mu\beta pk\rangle
&=&
  p_\mu\, \langle \beta\alpha pk\rangle
 -p^2\, \langle \mu\beta\alpha k \rangle
 +p.k\, \langle \mu\beta\alpha p\rangle\,.
\label{eq:redundant_terms_difference}
\end{eqnarray}
Here, $k_\beta$ and $k_\alpha$ are replaced effectively by $p_\beta$
and $-p_\alpha$, which is guaranteed by the divergence-free condition
(\ref{eq:divergence_free}) of the wave tensors. \s

The traceless, totally symmetric and divergence-free wave tensors
exclude any $g_{\alpha_i\alpha_j}$ and $g_{\beta_i\beta_j}$ terms and any
$\epsilon_{\alpha_i\alpha_j\cdots}$ and $\epsilon_{\beta_i\beta_j\cdots}$
terms. Then, the general form of each of the tensors
$F^a_{\beta,\alpha}(p,k)$ and $\bar{F}^a_{\alpha,\beta}(p,k)$ can be written
in terms of mutually independent $s+1$ parity-even and $s$ parity-odd parts
as
\begin{eqnarray}
 F^a_{\beta_1\cdots\beta_{s},\alpha_1\cdots\alpha_{s}}(p,k)
&=& F^a_1(k^2)\, g_{\beta_1\alpha_1}\cdots g_{\beta_{s}\alpha_{s}}
        + F^a_2(k^2)\, p_{\beta_1} p_{\alpha_1} g_{\beta_2\alpha_2}\cdots g_{\beta_{s}\alpha_{s}}\nonumber\\
&& + \,\cdots\,
        +F^a_{s+1}(k^2)\, p_{\beta_1} p_{\alpha_1} p_{\beta_2} p_{\alpha_2}\cdots
        p_{\beta_{s}} p_{\alpha_{s}}\nonumber \\[1mm]
&& + F'^a_1(k^2)\, \langle \beta_1\alpha_1 pk\rangle\, g_{\beta_2\alpha_2}
         \cdots g_{\beta_{s}\alpha_{s}} \nonumber\\
&& + F'^a_2(k^2)\, \langle \beta_1\alpha_1 pk\rangle\, p_{\beta_2} p_{\alpha_2}
          g_{\beta_3\alpha_3}
         \cdots g_{\beta_{s}\alpha_{s}}\nonumber\\
&& + \, \cdots\,
     + F'^a_{s}(k^2)\, \langle \beta_1\alpha_1 pk\rangle\,
     p_{\beta_2} p_{\alpha_2} p_{\beta_3} p_{\alpha_3}\cdots
     p_{\beta_{s}} p_{\alpha_{s}}\,,
\label{eq:f_tensor_decomposition_x2_x1}\\
 \bar{F}^a_{\alpha_1\cdots\alpha_{s},\beta_1\cdots\beta_{s}}(p,k)
&=& F^a_{\beta_1\cdots\beta_{s},\alpha_1\cdots\alpha_{s}}(p,k)\,
    [\, F^a_i \to \bar{F}^a_i\ \ \mbox{and}\ \
    F'^a_j \to \bar{F}'^a_j\,]\,,
\label{eq:f_tensor_decomposition_x1_x2}
\end{eqnarray}
with $a=1,2$, $i=1,\cdots\, s+1$ and $j=1,\cdots\, s$.
Similarily, the tensors $G$ and $\bar{G}$ in
Eqs.$\,$(\ref{eq:identical_spin_general_decomposition_x2_x1})
and (\ref{eq:identical_spin_general_decomposition_x1_x2}) can be
written in terms of $s$ independent parts as
\begin{eqnarray}
 G^b_{\beta_2\cdots\beta_{s},\alpha_2\cdots\alpha_{s}}(p,k)
&=& G^b_1(k^2)\, g_{\beta_2\alpha_2}\cdots g_{\beta_{s}\alpha_{s}}
        + G^b_2(k^2)\, p_{\beta_2} p_{\alpha_2} g_{\beta_3\alpha_3}\cdots g_{\beta_{s}\alpha_{s}}\nonumber\\
&& + \,\cdots\,
        +G^b_{s}(k^2)\, p_{\beta_2} p_{\alpha_2} p_{\beta_3} p_{\alpha_3}\cdots
        p_{\beta_{s}} p_{\alpha_{s}}\,,
\label{g_general_decomposition_x2_x1}\\
\bar{G}^b_{\alpha_2\cdots\alpha_{s},\beta_2\cdots\beta_{s}}(p,k)
&=& G^b_{\beta_2\cdots\beta_{s},\alpha_2\cdots\alpha_{s}}(p,k)\,
   [\,G^b_i \to \bar{G}^b_i\,]\,,
\label{g_general_decomposition_x1_x2}
\end{eqnarray}
for each of $b=1,2,3,4$.
It is important to note that substituting the pair $p_{\beta_j}p_{\alpha_j}$ or $g_{\beta_j\alpha_j}$ by the symmetric pair $\langle \beta_j\alpha_j pk\rangle$
with $j\geq 2$ does not introduce any new factor. One immediate consequence
of particular importance is that {\it in the identical spin case
with $s_2=s_1=s$ there are in general $8s+2$ independent form factors
for each of the $X_2\to X_1$ and $X_1\to X_2$ transition vertices of
two spin-$s$ on-shell particles $X_2$ and $X_1$. If the transversality
condition is valid, the number of independent terms reduces to
$6s+1=(8s+2)-(2s+1)$.}
Particularly, for the spin $s=1$, there are $6\times 1+1=7$
independent terms as pointed out through the general and comprehensive
study of the non-Abelian trilinear $WW\gamma$ and $WWZ$ couplings in
Ref.$\,$\cite{Hagiwara:1986vm}.   \s

\subsection{Different spin case: \boldmath{$s_2\neq s_1$}}
\label{subsec:different_spin_case}

For the sake of convenient discussion on the different spin case, the inequality
of $s_2 > s_1$ is assumed without any loss of generality. Then the vertex tensors
$\Gamma(p,k)$ and $\bar{\Gamma}(p,k)$ for the $X_2\to X_1$ and $X_1\to X_2$ vector-current
transitions in Eqs.$\,$(\ref{eq:vector_vertex_two_majorana_bosons_x2_x1})
and (\ref{eq:vector_vertex_two_majorana_bosons_x1_x2}) can be written as
\begin{eqnarray}
 \Gamma_{\mu; \beta_1\cdots\beta_{s_1},\alpha_1\cdots\alpha_{s_2}}(p,k)
&=&  p_\mu\, F^1_{\beta_1\cdots\beta_{s_1},\alpha_1\cdots\alpha_{s_2}}(p,k)
    + k_\mu\, F^2_{\beta_1\cdots\beta_{s_1},\alpha_1\cdots\alpha_{s_2}}(p,k)
    \nonumber\\
&+& (g_{\mu\beta_1} p_{\alpha_1} +  g_{\mu\alpha_1} p_{\beta_1})\,
     G^1_{\beta_2\cdots\beta_{s_1},\alpha_2\cdots\alpha_{s_2}}(p,k)
    \nonumber\\
&+& \langle \mu\beta_1\alpha_1 k\rangle\,
     G^2_{\beta_2\cdots\beta_{s_1},\alpha_2\cdots\alpha_{s_2}}(p,k)
    \nonumber\\
&+& (g_{\mu\beta_1}p_{\alpha_1}-g_{\mu\alpha_1}p_{\beta_1})\,
     G^3_{\beta_2\cdots\beta_{s_1},\alpha_2\cdots\alpha_{s_2}}(p,k)
    \nonumber\\
&+& \langle \mu\beta_1\alpha_1 p\rangle \,
     G^4_{\beta_2\cdots\beta_{s_1},\alpha_2\cdots\alpha_{s_2}}(p,k)
    \nonumber\\
&+& g_{\mu\alpha_1}\, T_1(k^2)\, g_{\beta_1\alpha_2}\cdots g_{\beta_{s_1}\alpha_{s_1+1}}
    \, p_{\alpha_{s_1+2}}\cdots p_{\alpha_{s_2}}
    \nonumber\\
&+& \langle \mu\alpha_1 pk\rangle\,  T_2(k^2)\, g_{\beta_1\alpha_2}
    \cdots g_{\beta_{s_1}\alpha_{s_1+1}}
    \, p_{\alpha_{s_1+2}}\cdots p_{\alpha_{s_2}}\,,
\label{eq:different_spin_general_decomposition_x2_x1}\\
\bar{\Gamma}_{\mu; \alpha_1\cdots\alpha_{s_2},\beta_1\cdots\beta_{s_1}}(p,-k)
&=& \Gamma_{\mu; \beta_1\cdots\beta_{s_1},\alpha_1\cdots\alpha_{s_2}}(p,k)\,
   [\, F^a\to \bar{F}^a,\, G^b \to \bar{G}^b,\, T_c \to \bar{T}_c\,]\,,
\label{eq:different_spin_general_decomposition_x1_x2}
\end{eqnarray}
with $a=1,2$, $b=1,2,3,4$ and $c=1,2$. In passing, we note again that the totally
symmetric wave tensors to be coupled to the vertices guarantee the {\it automatic
symmetrization of all the terms under any $\alpha$-index and/or $\beta$-index
permutations.}
It is crucial to note that, compared to the identical spin case, there exist
two additional form factors in the different spin case, the last two
terms in each of Eqs.$\,$(\ref{eq:different_spin_general_decomposition_x2_x1})
and (\ref{eq:different_spin_general_decomposition_x1_x2}).
With $s_2> s_1$, the tensors $F$ and $\bar{F}$ can be written in a factorized
form as
\begin{eqnarray}
  F^a_{\beta_1\cdots\beta_{s_1},\alpha_1\cdots\alpha_{s_2}}(p,k)
& =& F^a_{\beta_1\cdots\beta_{s_1},\alpha_1\cdots\alpha_{s_1}}(p,k)\,
     p_{\alpha_{s_1+1}}\cdots p_{\alpha_{s_2}} \,,
\label{eq:p_factorized_form_x2_x1}\\
 \bar{F}^a_{\alpha_1\cdots\alpha_{s_2},\beta_1\cdots\beta_{s_1}}(p,k)
& =& \bar{F}^a_{\alpha_1\cdots\alpha_{s_1},\beta_1\cdots\beta_{s_1}}(p,k)\,
      p_{\alpha_{s_1+1}} \cdots p_{\alpha_{s_2}} \,,
\label{eq:p_factorized_form_x1_x2}
\end{eqnarray}
with $a=1,2$
where each of the tensors $F^a_{\beta_1\cdots\beta_{s_1},\alpha_1\cdots\alpha_{s_1}}
(p,k)$ and $ \bar{F}^a_{\alpha_1\cdots\alpha_{s_1},\beta_1\cdots\beta_{s_1}}(p,k)$
takes the same form as the expression in each of Eqs.$\,$(\ref{eq:f_tensor_decomposition_x2_x1}) and
(\ref{eq:f_tensor_decomposition_x1_x2}) with the replacement of $s$ by $s_1$
in the $s_2> s_1$ case, i.e. each of the tensors consists of
mutually independent $s_1+1$ parity-even and $s_1$ parity-odd parts.
Similarily, each of the tensors $G$ and $\bar{G}$ in
Eqs.$\,$(\ref{eq:different_spin_general_decomposition_x2_x1}) and
(\ref{eq:different_spin_general_decomposition_x1_x2}) also can be factorized
as
\begin{eqnarray}
    G^b_{\beta_2\cdots\beta_{s_1},\alpha_2\cdots\alpha_{s_2}}(p,k)
&=& G^b_{\beta_2\cdots\beta_{s_1},\alpha_2\cdots\alpha_{s_1}}(p,k)\,
    p_{\alpha_{s_1+1}} \cdots p_{\alpha_{s_2}}\,,
\label{eq:q_factorized_form_x2_x1}\\
\bar{G}^b_{\alpha_2\cdots\alpha_{s_2},\beta_2\cdots\beta_{s_1}}(p,k)
&=& \bar{G}^b_{\alpha_2\cdots\alpha_{s_1},\beta_1\cdots\beta_{s_1}}(p,k)\,
    p_{\alpha_{s_1+1}} \cdots p_{\alpha_{s_2}} \,,
\label{eq:q_factorized_form_x1_x2}
\end{eqnarray}
with $b=1,2,3,4$ and each of the tensors $G^b$ and $\bar{G}^b$ consisting
of the $s_1$ independent parts as the expression in each of
Eqs.$\,$(\ref{g_general_decomposition_x2_x1}) and
(\ref{g_general_decomposition_x1_x2}) with the replacement of $s$ by $s_1$.
Consequently, {\it in the different spin case of $s_2\neq s_1$ there are
in general $8s+4$ independent form factors with $s={\rm min}(s_1, s_2)$ for
each of the $X_2\to X_1$ and $X_1\to X_2$ transition vertices of
a spin-$s_2$ on-shell particle $X_2$ and a spin-$s_1$
on-shell particle $X_1$. If the transversality condition is valid,
then the number of independent terms reduces to $6s+3=(8s+4)-(2s+1)$.}  \s

\subsection{Hermiticity and Majorana condition}
\label{subsec:hermiticity_majorana_condition}

The results presented in the previous subsections are applicable irrespective
of whether the particles $X_2$ and $X_1$ are charged or neutral. If any absorptive
parts are ignored and the particles are charge self-conjugate, i.e.
Majorana particles, then the vertex structure is strongly restricted.\s

Firstly, if any absorptive parts are ignored, i.e. the effective
Lagrangian, which is Hermitian, is used for constructing the $X_2\to X_1$
and $X_1\to X_2$ transition vector vertices, the following Hermiticity relation holds:
\begin{eqnarray}
  \bar{\Gamma}_{\mu;\alpha,\beta}(p,k)
\, =\,
  \Gamma^*_{\mu;\beta,\alpha}(p,k)\,.
\label{eq:hermitian_lagrangian}
\end{eqnarray}
Independently of whether the particles are charged or neutral,
the Hermiticity relation (\ref{eq:hermitian_lagrangian}) leads to the relations
for all the form factors as
\begin{eqnarray}
    \bar{F}^a_i(k^2)
&=& F^{a*}_i(k^2)\,,\\
    \bar{F}'^a_j(k^2)
&=& F'^{a*}_j(k^2)\,, \\
    \bar{G}^b_j(k^2)
&=& G^{b*}_j(k^2)\,, \\
    \bar{T}_{a}(k^2)
&=& T^*_{a}(k^2)\,,
\label{eq:hermiticity_form_factor_relation}
\end{eqnarray}
where $a=1,2$, $b=1,2,3,4$, $i=1,\cdots, s+1$, and $j=1,\cdots, s$ with
$s={\rm min}(s_2,s_1)$, so that the $X_1\to X_2$ transition vertex is
fixed once the $X_2\to X_1$ transition vertex is given. \s

Secondly, if the particles, $X_2$ and $X_1$, are not only neutral but also
charge self-conjugate, i.e. Majorana bosons, the crossing symmetry gives
an additional condition
\begin{eqnarray}
  \bar{\Gamma}_{\mu; \alpha,\beta} (p,k)
\, =\, \Gamma_{\mu; \beta,\alpha} (-p,-k)\,.
\label{eq:self_conjugate_relation}
\end{eqnarray}
Together with the Hermiticity condition (\ref{eq:hermitian_lagrangian}),
this charge self-conjugation or Majorana relation
(\ref{eq:self_conjugate_relation}) leads to the condition for the $X_2\to X_1$
transition (which will be called the Hermiticity-Majorana (HM) condition
in the following)
\begin{eqnarray}
  \Gamma_{\mu; \beta,\alpha} (p,k)
\, =\, \Gamma^*_{\mu; \beta, \alpha} (-p,-k)\,,
\label{eq:hermiticity_majorana__relation}
\end{eqnarray}
that is valid independently of whether the spacetime discrete
symmetries are conserved or not. It is straightforward to check
the following relations of all the form factors
\begin{eqnarray}
  F^{a}_i(k^2)
&=& - \eta_{21} \, F^{a*}_i(k^2) \,,\\
  F'^{a}_j(k^2)
&=& - \eta_{21} \, F'^{a*}_j(k^2) \,, \\
  G^{b}_j(k^2)
&=& - \eta_{21}\, G^{b*}_j(k^2)\,,\\
   T_{a}(k^2)
&=& - \eta_{21}\, T^*_{a}(k^2)\,,
\label{eq:hermiticity_majorana_form_factor_relation}
\end{eqnarray}
with $a=1,2$, $b=1,2,3,4$, $i=1,\cdots, s+1$, and $j=1,\cdots, s$,
and with a spin-dependent phase factor $\eta_{21}=(-1)^{s_2-s_1}$.
Therefore, the HM condition (\ref{eq:hermiticity_majorana__relation})
leads to the following selection rules:
\begin{itemize}
\item In the different spin case, for the even (odd) spin difference case
       with $\eta_{21}=+1(-1)$, all the $8s+4$ form factors are purely
       imaginary (purely real).
\item In contrast, in the identical spin case always with $\eta_{21}=+1$,
       the form factors, $T_{1,2}(k^2)$, are absent. As a result,
       all the $8s+2$ form  factors are purely imaginary.
\end{itemize}
These selection rules will be demonstrated explicitly by studying
the two-body decay $X_2\to V X_1$ with $V$, collectively denoting
an on-shell or off-shell vector boson for four spin combinations of
$(s_2, s_1)=(0, 0),\, (0, 1),\, (1, 0)$ and $(1,1)$.\s

If two Majorana bosons, $X_2$ and $X_1$, are identical, i.e.  $s_2=s_1$ and
$m_2=m_1$, Bose symmetry carries a further requirement that
the vertex tensor $\Gamma_{\mu;\beta,\alpha}(p,k)$ be symmetric under the
interchange of indices and momenta as $\beta\, \leftrightarrow\,
\alpha$ and $p_1\, \leftrightarrow\, -p_2$, resulting in the replacements,
$p\, \rightarrow\, -p$ and $k\, \rightarrow\, k$.
In this case, all of the $4s+1$ form factors, $F^{1}_i(k^2)$, $F'^{1}_j(k^2)$
and $G^{1,2}_j(k^2)$, are vanishing and {\it only the $4s+1$ form factors,
$F^{2}_i(k^2)$, $F'^{2}_j(k^2)$ and $G^{3,4}_j(k^2)$ can survive.} Consequently,
we have the following selection rules for the vertex of two identical Majorana
bosons worked in detail previously:
\begin{itemize}
\item Two identical Majorana spin-zero scalars do not couple to
      any on-shell vector boson or any conserved vector current at all,
      as the $F^{2}_i(k^2)$ and $F'^{2}_j(k^2)$ terms do not contribute and
      the $G^{3,4}_j(k^2)$ terms exist only for $s\geq 1$. It corresponds to
      the statement that a Majorana scalar cannot have any electromagnetic
      form factors.
\item For the spin $s\geq 1$, every term proportional to $p_\mu$ is
      forbidden, implying that an integer-spin Majorana particle cannot
      have any static electromagnetic moments.
\item All the $2s$ surviving terms are of the so-called anapole type, i.e.
      they simply give rise to a contact interaction.
\end{itemize}
All these characteristics are consistent with those derived and discussed
in detail in Refs.$\,$\cite{Boudjema:1988zs,Boudjema:1990st}.\s

If CP symmetry is also preserved in the $X_2\leftrightarrow X_1$ vector-current
transitions, then the following relation combined with the Majorana condition
is satisfied:
\begin{eqnarray}
\Gamma_{\mu; \beta,\alpha} (p,k)
\, =\, -n^*_2 n_1\, \Gamma^P_{\mu; \beta,\alpha} (p,k)\,,
\label{eq:cp_hermiticity_majorana__relation}
\end{eqnarray}
with the normalities defined to be $n_{1,2}=\eta_{1,2}\, (-1)^{s_{1,2}}$ in
terms of the intrinsic CP parities $\eta_{1,2}$ of the particles, $X_{1,2}$,
and under the assumption that the intrinsic CP parity of $V$ is even.
The superscript $P$ implying that the sign of every term involving
a totally antisymmetric Levi-Civita tensor needs to be flipped.
Consequently, CP invariance leads to the following
selection rules. {\it In the different spin case, with the same normality
of $n_2=n_1$, only the $4s+1$ parity-odd form factors, $F'^{1,2}_j(k^2)$,
$G^2_j(k^2)$, $G^4_j(k^2)$ and $T_2(k^2)$, survive,
and, with the opposite normality of $n_2=-n_1$, the other $4s+3$ parity-even
form factors, $F^{1,2}_i(k^2)$, $G^1_j(k^2)$, $G^3_j(k^2)$ and $T_1(k^2)$, survive.
In the identical spin case, as  the form factors $T_{1,2}(k^2)$ do not appear,
the number of independent terms reduce to $4s$ in the same normality case
and $4s+2$ in the opposite normality, while if the transversality condition
is valid, they further reduce to $3s$ and $3s+1$ in the same and opposite
normality cases, respectively.}\s

\subsection{The case with \boldmath{$V=\gamma$}}
\label{subsec:on-shell_photon}

When the vector boson $V$ is an on-shell massless photon $\gamma$,
the photon wave function and momentum should satisfy the on-shell
conditions
\begin{eqnarray}
   k \cdot \epsilon_\gamma (k,\lambda)=0
\quad \mbox{and} \quad
   k^2 = 0\,,
\label{eq:photon_on-shell_conditions}
\end{eqnarray}
with the $\gamma$ helicity $\lambda=\pm 1=\pm$. Imposing the on-shell
conditions (\ref{eq:photon_on-shell_conditions}) casts the triple
vertex $\Gamma$ into the reduced form
\begin{eqnarray}
  \Gamma_{\mu;\beta_1\cdots\beta_{s_1},\alpha_1\cdots\alpha_{s_2}}(p,k)
&=& (g_{{_\bot} \mu\beta_1}p_{\alpha_1}-g_{{_\bot}\mu\alpha_1}p_{\beta_1})\,
    G^1_{\beta_2\cdots\beta_{s_1},\alpha_2\cdots\alpha_{s_2}}(p,k)
   \nonumber\\
&+& (g_{{_\bot}\mu\beta_1}p_{\alpha_1}+g_{{_\bot}\mu\alpha_1}p_{\beta_1})\,
     G^2_{\beta_2\cdots\beta_{s_1},\alpha_1\cdots\alpha_{s_2}}(p,k)
    \nonumber\\
&+& \langle \mu\beta_1\alpha_1k\rangle\,
     G^3_{\beta_2\cdots\beta_{s_1},\alpha_1\cdots\alpha_{s_2}}(p,k)
    \nonumber\\
&+& \langle \mu\beta_1\alpha_1 p\rangle_{_\bot}\,
     G^4_{\beta_2\cdots\beta_{s_1},\alpha_1\cdots\alpha_{s_2}}(p,k)
    \nonumber\\
&+& g_{{_\bot}\mu\alpha_1}\, T_1(k^2)\,
    g_{\beta_1\alpha_2}\cdots g_{\beta_{s_1}\alpha_{s_1+1}}
    \, p_{\alpha_{s_1+2}}\cdots p_{\alpha_{s_2}}
    \nonumber\\
&+& \langle \mu\alpha_1 pk\rangle\,  T_2(k^2)\, g_{\beta_1\alpha_2}
    \cdots g_{\beta_{s_1}\alpha_{s_1+1}}
    \, p_{\alpha_{s_1+2}}\cdots p_{\alpha_{s_2}}\,,
\label{eq:general_x2x1_photon_coupling}
\end{eqnarray}
with $g_{{_\bot}\mu\rho}=g_{\mu\rho}-p_\mu k_\rho/p\cdot k $ for any four-vector
index $\rho$, and
$\langle \mu\beta\alpha p\rangle_{_\bot}
=\langle \mu\beta\alpha p\rangle +
\langle \beta\alpha pk\rangle\, p_\mu/p\cdot k$, both of which are
orthogonal to $k_\mu$. Consequently, {\it in the different and identical
spin cases, there are $4s+2$ and $4s$ independent form factors with
$s={\rm min}(s_2,s_1)$, respectively, as the form factors $T_{1,2}(k^2)$
do not appear in the identical spin case.} In passing, we note that
the $X_2X_1\gamma$ vertex
structure for the identical spin case of $s_2=s_1=1$ has $4\times 1=4$ independent terms as pointed out and studied in detail in
Refs.$\,$\cite{Renard:1981es,Hagiwara:1986vm,Choudhury:1994nt,
Ananthanarayan:2004eb,Rahaman:2016pqj,Rahaman:2017qql,Rahaman:2020jll,Choi:1994nv}. \s

\subsection{The case with \boldmath{$X_1=V=\gamma$}}
\label{subsec:two_on-shell_photons}

As a special case, let us consider the decay of a massive integer-spin
Majorana boson into two photons, $X_2\to \gamma\gamma$, corresponding
to taking $X_1=V=\gamma$. Imposing the on-shell conditions
\begin{eqnarray}
&& p_1\cdot \epsilon_1(p_1,\lambda_1)=0\quad \mbox{and} \quad
   p^2_1=0\,, \\[1mm]
&& k\cdot \epsilon_\gamma(k, \lambda)=0\quad\, \ \ \, \, \mbox{and} \quad
  k^2=0\,,
\label{eq:two_photons}
\end{eqnarray}
with $\lambda_1, \lambda=\pm 1=\pm$, and performing the Bose symmetrization
of two identical photon states allow us to write the general $X_2\gamma\gamma$
vertex in a greatly-simplified form as
\begin{eqnarray}
   \Gamma^{X_2\gamma\gamma}_{\mu\beta_1;\alpha_1\cdots\alpha_{s_2}}(p_2,q)
&=& \eta_+
    Y^+_1\,\, g_{{_\bot} \mu\beta_1}\,\, q_{\alpha_1}\cdots q_{\alpha_{s_2}}
    \nonumber\\
&+& \eta_+\,
    Y^+_2\,\, \langle \mu\beta_1 p_2 q\rangle\,\,
               q_{\alpha_1}\cdots q_{\alpha_{s_2}}
    \nonumber\\
&+& \eta_+\,
    Y^+_3\,\, \left[\, g_{{_\bot}\mu\alpha_1} g_{{_\bot}\beta_1\alpha_2}
                           + g_{{_\bot}\beta_1\alpha_1} g_{{_\bot}\mu\alpha_2}\,
                           \right]\,\,
                    q_{\alpha_3}\cdots q_{\alpha_{s_2}}
    \nonumber\\
&+& \eta_-\,
    Y^-_1\,\, \left[\, g_{_{\bot}\mu\alpha_1}\langle \beta_1\alpha_2 p_2 q\rangle
                           + g_{_{\bot}\beta_1\alpha_1}\langle \mu\alpha_2 p_2 q\rangle
                          \, \right]\,\,
                    q_{\alpha_3}\cdots q_{\alpha_{s_2}}\,,
\label{eq:general_x2_two_photon_coupling}
\end{eqnarray}
with the projection factors, $\eta_\pm =[1\pm(-1)^{s_2}]/2$, and
two momentum combinations, $p_2=k+p_1$ and $q=k-p_1$, which are symmetric
and antisymmetric under the interchange of two photons, i.e. $k\leftrightarrow p_1$
and $\mu\leftrightarrow \beta_1$, respectively.
For the sake of notation, the following orthogonal tensors are introduced,
\begin{eqnarray}
    g_{_{\bot}\mu\beta_1}
&=& g_{\mu\beta_1}-p_{1\mu}k_{\beta_1}/p_1\cdot k\,,
    \\
    g_{_{\bot}\mu\alpha_i}
&=& g_{\mu\alpha_i}-p_{1\mu} k_{\alpha_i}/p_1\cdot k\,,
    \\
    g_{_{\bot}\beta_1\alpha_i}
&=& g_{\beta_1\alpha_i} - k_{\beta_1}p_{1\alpha_i}/p_1\cdot k\,,
\label{eq:orthgonal_tensors}
\end{eqnarray}
with $i=1,2$.
The parity of each term in Eq.$\,$(\ref{eq:general_x2_two_photon_coupling})
is determined according to whether its sign
flips or not when the sign of $q$ is changed. It is now straightforward
to derive the following selection rules from the expression
(\ref{eq:general_x2_two_photon_coupling}) of the $X_2\gamma\gamma$ vertex,
\begin{itemize}
\item The parity-even $Y^+_{1}$ and parity-odd $Y^+_{2}$
      terms survive for $s_2=0, 2, 4$, etc.
\item The parity-even $Y^+_{3}$ term survives for $s_2=2,4$, etc.
\item The parity-even $Y^-_1$ term survives for $s_2=3,5$, etc.
\end{itemize}
Combining these results together we can count the number
$n$ of possible even/odd-parity ($P=\pm $) states of the
two-photon system for a given $X_2$ integer-spin $s_2$.
The selection rules can be summarized collectively with the
compact notation $n [s_2]^{P}$ as
\begin{eqnarray}
{\large n\, [s_2]^P} \ \ = \ \ 1\,[0]^+\,, \ \ 2\,[2k]^+\,,\ \
             1\,[2k+1]^+\,, \ \  1\,[0]^-\,,\ \ 1\, [2k]^-\,,
\label{eq:landau_yang_theorem}
\end{eqnarray}
with the positive integer $k=1,2,\cdots$, as worked out independently
by Landau~\cite{Landau:1948kw} and
Yang~\cite{Yang:1950rg}. One immediate consequence of the so-called
Landau-Yang theorem is that {\it any massive on-shell spin-1 particle
with $s_2=1$ cannot decay into two on-shell photons.} Accordingly, as
the resonance with mass about 125 GeV discovered at the LHC has been
observed to decay into two on-shell
photons~\cite{Aad:2012tfa,Chatrchyan:2012ufa}, its spin cannot be 1. \s

\section{Decay helicity amplitudes}
\label{sec:decay_helicity_amplitudes}

Complementary to the covariant formalism used in the previous section,
the helicity formalism~\cite{Wick:1962zz,Leader:2001gr,Chung:1971ri} is one
of the most effective tools for discussing the two-body decay of an on-shell
Majorana particle $X_2$ of mass $m_2$ and spin $s_2$ into an on-shell or
off-shell vector boson $V$ of mass $m$ ($=m_V$ for an on-shell $V$) and
an on-shell Majorana particle $X_1$ of mass $m_1$ and spin $s_1$, irrespective
of whether the spins $s_2$ and $s_1$ are integer or half-integer. \s

For the sake of a transparent analytic analysis, we describe the
two-body decay, $X_2\to V X_1$,
\begin{eqnarray}
X_2(p_2,\lambda_2)
   \ \ \rightarrow\ \
V(k,\lambda)\, +\, X_1(p_1,\lambda_1)\,,
\end{eqnarray}
in the $X_2$ rest frame ($X_2$RF) and the two-body leptonic decay of an
on-shell or off-shell vector boson, $V\to \ell^-\ell^+$,
\begin{eqnarray}
V(k^\prime,\lambda^\prime)
   \ \ \rightarrow\ \
\ell^-(q_-,\sigma_-)\, + \ell^+(q_+,\sigma_+)\,,
\end{eqnarray}
in the $V$ rest frame ($V$RF) directly reconstructible event by event
by measuring the momenta of two charged leptons with $\ell=e$ or $\mu$
with good precision. The momentum and helicity of each particle are shown in
parenthesis with the primed momentum $k^\prime$ referring to the momentum
in the $V$RF. One crucial point to be taken into account in combining the
two sequential decay amplitudes is that the $V$-boson polarization state
in the $X_2$RF is in general different from that in the $V$RF directly
reconstructed in the laboratory frame (LAB). \s

Before going into a detailed description of the angular correlations
in Section~\ref{sec:correlated_angular_distributions}, we study some
general restrictions on the decay helicity amplitudes
due to CP invariance and the Majorana condition that the particles,
$X_2$ and $X_1$ are their own antiparticles.\s

\subsection{Correlated decay helicity amplitudes}
\label{subsec:correlated_decay_helicity_amplitudes}

In general, a virtual vector boson $V$ in its rest frame has a zeroth
scalar component as well as three spin-1 space components. However,
the scalar component does not contribute to the decay amplitudes meaningfully,
if the virtual boson couples to a nearly conserved vector current like
the SM $\gamma$ and $Z$ vector currents of the $e$ and $\mu$ leptons
due to negligible $e$ and $\mu$ masses. In this light, the transversality
condition is assumed to be valid with very good approximation in
the following. Then, the $V$ invariant-mass dependent decay
helicity amplitude can be decomposed in terms of the polar and azimuthal
angles, $\theta$ and $\phi$, of the momentum direction of the boson
$V$ in the $X_2$RF in the Wick convention as
\begin{eqnarray}
  {\cal M}^{X_2\to V X_1}_{\lambda_2;\lambda,\lambda_1}(m; \theta,\phi)
= {\cal C}_{\lambda,\lambda_1}(m) \,
  d^{s_2}_{\lambda_2,\lambda-\lambda_1}(\theta)\,
  e^{i\lambda_2\phi}
  \quad\mbox{with}\quad
  |\lambda-\lambda_1|\leq s_2\,,
\label{eq:x2_vx1_helicity_amplitude}
\end{eqnarray}
with $\lambda_2=-s_2,\cdots, s_2$, $\lambda=\pm 1,\, 0=\pm,\,0 $ and
$\lambda_1=-s_1,\cdots, s_1$ with the constraint $|\lambda-\lambda_1|\leq s_2$.
The reduced helicity amplitudes ${\cal C}_{\lambda,\lambda_1}(m)$ do not depend
on any $X_2$ helicity $\lambda_2$ due to rotational invariance. The polar-angle
dependent function $d^{s_2}_{\lambda_2,\lambda-\lambda_1}(\theta)$ is a Wigner
$d$ function in the convention of Rose~\cite{merose2011}.\s

Based on the helicity-amplitude decomposition in
Eq.$\,$(\ref{eq:x2_vx1_helicity_amplitude}) and the restriction of
$|\lambda-\lambda_1|\leq s_2$ on the helicities, it is straightforward to
count the number of independent reduced helicity amplitudes {\it even
without knowing explicit forms of the reduced helicity amplitudes}.
In the identical
spin case of $s_2=s_1=s$, two maximal helicity-difference combinations
$(\lambda,\lambda_1)=(\pm 1, \mp s)$ among $(2 \times 1+1)\times (2s+1)=6s+3$
combinations of the $V$ and $X_1$ helicities are forbidden because of
the constraint $|\lambda-\lambda_1|\leq s$. As a result, {\it in the
identical spin case, the number of independent terms is $6s+1$, the same as
counted in the covariant description.} On the other hand, if $s_2> s_1$,
the constraint does not play any role so that the number of independent terms
is simply $3\times (2s_1+1) = 6s_1+3$. For $s_2<s_1$, the constraint plays a crucial
role in counting the number of degrees of freedom. For $\lambda=0$,
the $X_1$ helicity $\lambda_1$ can take $2s_2+1$ values from $-s_2$
to $s_2$, while for each of $\lambda = \pm 1$, it takes $2s_2+1$ values from
$-s_2\pm 1$ to $s_2\pm 1$. Therefore, the number of independent terms is
$3\times (2s_2+1)=6s_2+3$. Consequently, {\it in the different spin case,
the number of independent terms is $6s+3$ with $s={\rm min}(s_2, s_1)$, the
same as counted in the previous covariant description again.} \s

Because generally the momentum direction of the boson $V$ in the $X_2$RF
is different from that in the laboratory frame (LAB), the helicity amplitude
in Eq.$\,$(\ref{eq:x2_vx1_helicity_amplitude}) needs to be transformed
by a proper Wick helicity rotation~\cite{Choi:2019aig,Leader:2001gr}
for connecting the $V$ helicity state
in the $X_2$RF to that in the LAB with a so-called Wick helicity rotation
angle $\omega$ satisfying
\begin{eqnarray}
    \cos\omega
&=& \frac{\beta+\beta_2\cos\theta}{
          \sqrt{(1+\beta_2\beta\cos\theta)^2-(1-\beta^2_2)(1-\beta^2)}}\,,
\label{eq:cos_wick_helicity_rotation_angle}\\
    \sin\omega
&=& \frac{\sqrt{1-\beta^2}\, \beta_2\sin\theta}{
          \sqrt{(1+\beta_2\beta\cos\theta)^2-(1-\beta^2_2)(1-\beta^2)}}\,,
\label{eq:sin_wick_helicity_rotation_angle}
\end{eqnarray}
where $\beta_2$ and $\beta$ are the $X_2$ speed in the LAB and
the $V$ speed in the $X_2$RF, which are unambiguously determined
in terms of the $X_2$ energy in the LAB and the $X_{1,2}$ and $V$
masses. The resulting decay helicity amplitude to be directly coupled
with the $V$ decay helicity amplitude in the LAB reads\footnote{We do not
include another Wick helicity rotation connecting the $X_1$ helicity states
in the LAB and in the $X_2$RF because its effects on any distributions are
washed away completely with the summation over the $X_1$ helicities.}
\begin{eqnarray}
  {\cal A}_{\lambda_2;\lambda^\prime,\lambda_1}(\theta,\phi)
= \sum_{\lambda=\pm 1, 0}\, d^1_{\lambda^\prime,\lambda}(\omega)\,\,
  {\cal M}^{X_2\to V X_1}_{\lambda_2;\lambda,\lambda_1}(\theta,\phi)\,,
\label{eq:wick_helicity_rotated_x2_zx1_decay_helicity_mplitude}
\end{eqnarray}
It is important to note that the Wick helicity rotation angle $\omega$
along with the polar angle $\theta$ is determined event by event, although
it might not be possible to determine the azimuthal angle $\phi$
unambiguously. \s

Among various decay channels of the $V$ boson, if available, the leptonic
decays $V\to\ell^-\ell^+$, especially with $\ell=e$ and $\mu$, can provide
a very clean and powerful means for reconstructing the rest frame of the
boson $V$, independently of its production mechanisms, and
for extracting the information on $V$ polarization efficiently.
The helicity amplitude of the leptonic decay to be directly combined
with the $X_2\to VX_1$ helicity amplitude in
Eq.$\,$(\ref{eq:wick_helicity_rotated_x2_zx1_decay_helicity_mplitude})
can be written as
\begin{eqnarray}
  {\cal M}^{V\to \ell^-\ell^+}_{\lambda^\prime;\,\sigma_-,\sigma_+}
  (\theta_\ell,\phi_\ell)
= {\cal Z}_{\sigma_-,\sigma_+}(m)\,
  d^{1}_{\lambda^\prime,\sigma_--\sigma_+}(\theta_\ell)\,
  e^{i\lambda^\prime \phi_\ell}\,,
\label{eq:z_ll_helicity_amplitude}
\end{eqnarray}
in terms of the $\ell^-$ polar and azimuthal angles, $\theta_\ell$ and
$\phi_\ell$, in the $V$RF with the azimuthal angle which can be defined
with respect to the plane formed by the $V$ momentum direction and
an appropriately-chosen non-parallel direction fixed in the LAB. \s

\subsection{Discrete spacetime symmetries and Majorana condition}
\label{subsec:discrete_spacetime_symmetries_majorana_condition}

Even in transitions involving weak interactions, the decay processes
observe CP symmetry to a great extent while often violating P and C
symmetries significantly.
So we discuss the consequences of the CP symmetry among discrete spacetime
symmetries in the decay helicity amplitudes. For the decay processes
involving two Majorana particles $X_2$ and $X_1$, CP invariance leads
to the following relation for the reduced helicity amplitudes in Eq.$\,$(\ref{eq:x2_vx1_helicity_amplitude}) as
\begin{eqnarray}
    {\cal C}_{\lambda,\lambda_1}(m)
= - n^*_2 n_1 \,\, {\cal C}_{-\lambda,-\lambda_1}(m)\,,
\label{eq:x2_v_x1_cp_relation}
\end{eqnarray}
with the $X_2$ and $X_1$ normalities, $n_2=\eta_2 (-1)^{s_2}$ and
$n_1=\eta_1 (-1)^{s_1}$, in terms of the intrinsic CP parities,
$\eta_2$ and $\eta_1$, under the assumption that the normality of $V$
is $-1$ with $s_V=1$ and even CP parity like $\gamma$ and $Z$ in the SM.
Note that the CP symmetry test does not assume the absence of any absorptive
parts and rescattering effects. Certainly, the CP relation
(\ref{eq:x2_v_x1_cp_relation}) of the reduced helicity amplitudes is closely
related to the CP relation (\ref{eq:cp_hermiticity_majorana__relation})
of the triple vertex tensor.\s

Together with CPT invariance, the Majorana condition that both of the two
neutral particles $X_2$ and $X_1$ are their own antiparticles leads to the
relation for the decay helicity amplitudes in
Eq.$\,$(\ref{eq:x2_vx1_helicity_amplitude}),
\begin{eqnarray}
  {\cal C}_{\lambda,\lambda_1}(m)
= -\eta_{21}\, {\cal C}^*_{-\lambda,-\lambda_1}(m)\,,
\label{eq:x2_v_x1_majorana_condition}
\end{eqnarray}
with the sign factor $\eta_{21}=(-1)^{s_2-s_1}$
in the absence of any absorptive parts and rescattering effects. Certainly,
this HM relation (\ref{eq:x2_v_x1_majorana_condition}) of the reduced
helicity amplitudes reflects the equivalent HM relation
(\ref{eq:hermiticity_majorana__relation}) of the triple vertex tensors.\s

As a representative explicit set of the reduced helicity amplitudes,
the most general $X_2X_1V$ tensor couplings for two Majorana
particles, $X_2$ and $X_1$, of spin$\,\,\leq\, 1$ are listed in
Table~\ref{tab:general_tensor_x2x1v_couplings}. The same and opposite normality
cases are treated separately, although the analysis in the mixed normality case
proceeds as in the fixed normality case, since the most general vertex is
the sum of the same and opposite normality cases.
For the sake of notation, we use simple alphabetic notations, $a, b, c, d$ and
$\bar{a}, \bar{b}, \bar{c}, \bar{d}$ for denoting the independent form factors,
dependent generally on the
invariant mass $m$ of the on-shell or off-shell vector boson $V$.
Taking a specific numerical set of masses and couplings, we present
a few numerical analyses for probing
the spin and dynamical structure of the two-body decays $X_2\to V X_1$ directly
related to the general $X_2X_1V$ vertex in
Subsection~\ref{seubsec:numerical_investigations_correlated_distributions}. \s

\vskip 0.8cm

\begin{table}[H]
\centering
\begin{tabular}{||c||l|l|c||} \hhline{|t:=:t:===:t|}
  { \boldmath{$s_2, s_1$}}
& \multicolumn{1}{c|}{\bf \small $X_2X_1V$ Coupling}
& \multicolumn{1}{c|}{\bf \small Reduced helicity amplitudes}
& {\bf \small Threshold}
  \\
  \hhline{|:=:b:===:|}
\multicolumn{4}{||c||}{\color{blue} Same normality\,:
                       $n_2=n_1\, \Rightarrow\,\,
                        {\cal C}_{-\lambda,-\lambda_1}=-C_{\lambda,\lambda_1} $}
  \\
  \hhline{|:=:t:===:|}
&
&
&
  \\
  \rb{1.5ex}{\small $0,0$}
& \rb{1.5ex}{\phantom{+} --}
& \rb{1.5ex}{\ \ --}
& \rb{1.5ex}{--}
  \\
  \hhline{||-||---||}
&
& $\sts \ \ {\cal C}_{0,0}\,=\,0 $
& --
  \\
  \rb{1.5ex}{\small $0,1$}
& \rb{1.5ex}{\small $\phantom{+}  b_1\, \langle \mu\beta pk\rangle $}
& $\sts \ \ {\cal C}_{1,1}\,=\, \imath\, b_1 \kappa_*\, m^2_2$
& $\sts \kappa_* $
  \\
  \hhline{||-||---||}
&
& $\sts \ \ {\cal C}_{0,0}\,=\, 0 $
& --
  \\
  \rb{1.5ex}{\small $1,0$}
& \rb{1.5ex}{$\phantom{+} c_1\, \langle \mu\alpha pk\rangle $}
& $\sts \ \ {\cal C}_{1,0}\,=\,\imath\, c_1 \kappa_*\, m^2_2 $
& $\sts \kappa_* $
  \\
  \hhline{||-||---||}
&
& $\sts \ \ {\cal C}_{0,0}\, =\, 0 $
& --
  \\
&
& $\sts \ \ {\cal C}_{0,1}\,=\,-\imath\, d_1 (m^2_2-m^2_1)/m$
& $\sts 1$
 \\
& $ \phantom{+} d_1\, \langle \mu\beta\alpha p\rangle
              + d_2\, \langle \mu\beta\alpha k\rangle $
& $\sts \ \ \phantom{{\cal C}_{0,1}\,=\,}
             -\imath\, d_2 m +\imath\, d_3 \kappa^2_*\, m^4_2/m$
&
 \\
  \rb{1.5ex}{\small $1,1$}
&  $ + d_3\,\, p_\mu\, \langle \beta\alpha pk\rangle $
& $\sts \ \ {\cal C}_{1,0}\,=\,-\imath\, d_1(m^2_2+3m^2_1-m^2)/2m_1 $
& $\sts 1$
  \\
&
& $\sts \ \ \phantom{{\cal C}_{1,0}\,=\,}-\imath\, d_2(m^2_2-m^2_1-m^2)/2m_1 $
&
  \\
&
& $\sts \ \ {\cal C}_{1,1}\,=\, -\imath\, d_1 (3m^2_2+m^2_1-m^2)/2m_2$
& $\sts 1 $
  \\
&
& $\sts \ \ \phantom{{\cal C}_{1,1}\,=\,} -\imath\, d_2 (m^2_2-m^2_1+m^2)/2m_2$
&
  \\
  \hhline{|:=:b:===:|}
\multicolumn{4}{||c||}{\color{blue} Opposite normality\,:
                        $n_2=-n_1\, \Rightarrow\,\,
                        {\cal C}_{-\lambda,-\lambda_1}=C_{\lambda,\lambda_1} $}
  \\
  \hhline{|:=:t:===:|}
&
&
&
  \\
  \rb{1.5ex}{\small $0,0$}
& \rb{1.5ex}{$\phantom{+} \bar{a}_1\, p_\mu$}
& \rb{1.5ex}{$\sts \ \ {\cal C}_{0,0}\, =\, \bar{a}_1 \kappa_*\, m^2_2/m  $}
& \rb{1.5ex}{$\sts \kappa_* $}
  \\
  \hhline{||-||---||}
&
& $\sts \ \ {\cal C}_{0,0}\, =\, \bar{b}_1\, (m^2_2-m^2_1-m^2)/2m_1 m $
& $\sts 1$
  \\
&
& $\sts \ \ \phantom{{\cal C}_{0,0}\, =\,} +\bar{b}_2\, \kappa^2_*\, m^4_2/2 m_1 m $
&
  \\
  \rb{2.5ex}{\small $0,1$}
& \rb{2.5ex}{\small $\phantom{+}  \bar{b}_1\, g_{\mu\beta}
                          +\bar{b}_2\, p_\mu p_\beta$}
& $\sts \ \ {\cal C}_{1,1}\,=\, -\bar{b}_1 $
& $\sts 1$
  \\
  \hhline{||-||---||}
&
& $\sts \ \ {\cal C}_{0,0}\,=\, -\bar{c}_1\, (m^2_2-m^2_1+m^2)/2m_2 m $
& $\sts 1$
  \\
&
& $\sts \ \ \phantom{{\cal C}_{1,0}\,=\,} +\bar{c}_2\, \kappa^2_*\, m^3_2/2 m$
&
  \\
\rb{2.5ex}{\small $1,0$}
& \rb{2.5ex}{$\phantom{+} \bar{c}_1\, g_{\mu\alpha}
                         +\bar{c}_2\, p_\mu p_\alpha$}
& $\sts \ \ {\cal C}_{1,0}\, =\, -\bar{c}_1$
& $\sts 1$
  \\
  \hhline{||-||---||}
&
& $\sts \ \ {\cal C}_{0,0}\,=\, \bar{d}_1\, \kappa_*\, m_2(m^2_2+m^2_1-m^2)/2m_1 m $
& $\sts \kappa_*$
  \\
&
& $\sts \ \ \phantom{{\cal C}_{0,0}\,=\,}+\bar{d}_2\, \kappa_*\, m_2(m^2_1-m^2)/2m_1 m$
&
  \\
& $ \phantom{+} \bar{d}_1\, p_\mu g_{\beta\alpha}
              + \bar{d}_2\, (g_{\mu\beta} p_\alpha
                           +g_{\mu\alpha} p_\beta)$
& $\sts \ \ \phantom{{\cal C}_{0,0}\,=\,}+\bar{d}_3\, \kappa_*\, m_2 (m^2_2-m^2_1)/2 m_1 m$
&
  \\
  \rb{1.5ex}{\small $1,1$}
& $ + \bar{d}_3\, (g_{\mu\beta} p_\alpha
                  -g_{\mu\alpha} p_\beta)
    + \bar{d}_4\, p_\mu p_\beta p_\alpha $
& $\sts \ \ \phantom{{\cal C}_{0,0}\,=\,}+\bar{d}_4\, \kappa^3_*\, m^5_2/4m_1 m $
&
  \\
&
& $\sts \ \ {\cal C}_{0,1}\,=\,\bar{d}_1\, \kappa_*\, m^2_2/m $
& $\sts \kappa_*$
 \\
&
& $\sts \ \ {\cal C}_{1,0}\,=\,-(\bar{d}_2-\bar{d}_3)\, \kappa_*\, m^2_2/2m_1 $
& $\sts \kappa_*$
  \\
&
& $\sts \ \ {\cal C}_{1,1}\,=\,-(\bar{d}_2+\bar{d}_3)\, \kappa_*\, m_2/2 $
& $\sts \kappa_*$
  \\
  \hhline{|b:=:b:===:b|}
\end{tabular}
\vskip 0.5cm
\caption{\it The most general $X_2X_1V$ tensor couplings and the corresponding
reduced helicity amplitudes for $X_2$ and $X_1$ of spin~$\leq 1$.
Here $p=p_2+p_1$ and $k=p_2-p_1$, where $p_{2,1}$ and $m_{2,1}$ are
the momenta and masses of the on-shell particles, $X_2$ and $X_1$,
and $m=\sqrt{k^2}$. The kinematical factor $\kappa_*=\lambda^{1/2}(1, m^2_1/m^2_2,
m^2/m^2_2)$ related to the $V$ speed in the $X_2$RF is approximately
proportional to $\sqrt{(m_2-m_1)-m}$ just below the threshold.
Note that the reduced helicity amplitudes, ${\cal C}_{00}$, are vanishing
in the same normality case as these decay modes are forbidden due to
angular momentum conservation .}
\label{tab:general_tensor_x2x1v_couplings}
\end{table}
\section{Correlated invariant-mass and polar-angle distributions}
\label{sec:correlated_angular_distributions}

The fully-correlated decay amplitudes will be helpful for probing
the polarization phenomena through which the spin and dynamical structures
of the interaction vertices are decoded.
In this section, firstly we derive all the analytic expressions for the
correlated invariant-mass and polar-angle distributions for the two sequential
decays, $X_2\to V X_1$ and $V\to \ell^-\ell^+$ with $\ell=e$ and $\mu$, which
consist of two helicity-dependent parts. Secondly, we check all the analytic
results by analyzing  four different spin combinations of
$(s_2, s_1)=(0,0), (0,1), (1,0)$ and $(1,1)$ numerically in two sets of
masses, of which one set is for an off-shell $V$ and the other set for an
on-shell $V$. \s

\subsection{Analytic derivation of the correlated distributions}
\label{seubsec:analytic_derivation_correlated_distributions}

We assume that the decaying $X_2$ particle is unpolarized on average~\footnote{
As shown explicitly in Ref.$\,$\cite{Choi:2020spm}, the parity-odd polarizations
of the Majorana particle $X_2$ of any spin produced in the process
$e^-e^+\to X_2 X_1$ is indeed vanishing on average.}
but it may have a known energy profile. As pointed out before,
it is necessary to include a Wick helicity rotation for calculating the combined
helicity  amplitude of the sequential decay of two 2-body decays $X_2\to
V X_1$ and  $V\to\ell^-\ell^+$ with $\ell=e$ or $\mu$. Note that
the polar angle $\theta_\ell$ of the charged lepton $\ell^-$ in the decay
$V\to\ell^-\ell^+$ can be measured event by event and so the $\theta_\ell$
distribution can be determined unambiguously. Integrating the distribution over
the lepton azimuthal angle $\phi_\ell$, which is usually difficult to be
reconstructed, casts the $V$ leptonic-decay density matrix depending on
the reconstructible polar angle $\theta_\ell$ into a diagonal form
\begin{eqnarray}
  \rho^V_{\lambda^\prime,\lambda^\prime}(\theta_\ell)
= \frac{1}{4}\,{\rm diag}\left(1+\cos^2\theta_\ell
                          + 2A_\ell\, \cos\theta_\ell,\ \
                          2\sin^2\theta_\ell,\ \
                          1+\cos^2\theta_\ell - 2A_\ell\, \cos\theta_\ell\right)\,,
\end{eqnarray}
in the $(+1, 0, -1)$ basis with the parity-odd factor
$A_\ell=2v_\ell a_\ell/(v^2_\ell+a^2_\ell)$ in terms of the normalized $V\ell\ell$
vector and axial-vector couplings $v_\ell$ and $a_\ell$. Numerically, for $V=Z$,  $A_\ell\simeq -0.16$~\cite{Zyla:2020zbs}. Then, the correlated invariant-mass and
polar-angle distribution independent of the production mechanism reads
\begin{eqnarray}
   \frac{d\Gamma[X_2\to V X_1\to\ell^-\ell^+ X_1]}{
         d m\, d\cos\theta\, d\cos\theta_\ell}
&=& \frac{2 m^3}{(m^2-m^2_V)^2+m^2_V \Gamma^2_V}\,
    \frac{d{\cal D}[X_2\to V X_1\to\ell^-\ell^+ X_1]}{
          d\cos\theta\, d\cos\theta_\ell}\,,
\label{eq:correlated_invariant_mass_polar_angle_distributions}
\end{eqnarray}
where the correlated polar-angle distribution is
given by
\begin{eqnarray}
  \frac{d{\cal D}[X_2\to V X_1\to\ell^-\ell^+ X_1]}{
        d\cos\theta\, d\cos\theta_\ell}
&=& \frac{3 \kappa_*\, \Gamma[V\to\ell^-\ell^+](m)}{
        64 (2s_2+1) \pi^2 m_2\, m}
  \sum_{\lambda^\prime,\lambda,\lambda_1}
  [d^1_{\lambda^\prime,\lambda}(\omega)|^2\,
  |C_{\lambda,\lambda_1}(m)|^2\,
  \rho^V_{\lambda^\prime,\lambda^\prime}(\theta_\ell)\nonumber\\
&=& \frac{3 \Gamma[V\to\ell^-\ell^+](m_V)}{
        64 (2s_2+1) \pi^2 m_2\, m_V}\,\kappa_*\,
  \sum_{\lambda^\prime}\,
  W_{\lambda^\prime,\lambda^\prime}(m,\omega)\,
  \rho^V_{\lambda^\prime,\lambda^\prime}(\theta_\ell)\,,
\label{eq:correlated_polar_angle_distributions}
\end{eqnarray}
with $\lambda^\prime, \lambda=\pm 1, 0$ and $\lambda_1=-s_1,\cdots, s_1$
satisfying the constraint $|\lambda-\lambda_1|\leq s_2$ and with the
kinematical phase factor $\kappa_* =\lambda^{1/2}(1, m^2/m^2_2, m^2_1/m^2_2)$.
In the last expression, we have taken into account
the fact that $\Gamma[V\to \ell^-\ell^+](m)$ scales in proportion to the invariant mass
$m$, when the lepton masses are ignored.
For the sake of discussion, the so-called Wick distribution function (WDF) $W_{\lambda',\lambda'}(m,\omega)$ as defined in Ref.$\,$\cite{Choi:2019aig}
is introduced:
\begin{eqnarray}
  W_{\lambda^\prime,\lambda^\prime}(m,\omega)
= \sum_{\lambda,\lambda_1}\,
          [\, d^1_{\lambda^\prime,\lambda}(\omega)]^2\,
          |C_{\lambda,\lambda_1}(m)|^2\,,
\label{eq:extended_wick_distribution_function}
\end{eqnarray}
with the constraint $|\lambda-\lambda_1|\leq s_2$,
where $\omega$ is a function of not only $\cos\theta$ but also $m$
as can be checked with Eqs.$\,$(\ref{eq:cos_wick_helicity_rotation_angle})
and (\ref{eq:sin_wick_helicity_rotation_angle}). This WDF encodes the
information on the spin and dynamical structure
of the two-body decay $X_2\to V X_1$ fully.\s

If the mass difference $m_2-m_1$ is larger than the vector-boson mass $m_V$
and also the width $\Gamma_V$ is much smaller than the mass $m_V$, we can take
the narrow-width approximation (NWA),
\begin{eqnarray}
     \frac{2 m^3}{(m^2-m^2_V)^2+m^2_V\Gamma^2_V}\quad
\rightarrow\quad
     \pi\, \frac{m_V}{\Gamma_V}\, \delta(m-m_V)\,,
\label{eq:narrow_width_approximation}
\end{eqnarray}
and then the correlated polar-angle distribution and the total width are
given by
\begin{eqnarray}
  \frac{d\Gamma[X_2\to V X_1\to\ell^-\ell^+ X_1]}{
        d\cos\theta\, d\cos\theta_\ell}
&=& \frac{3\, {\rm Br}[V\to\ell^-\ell^+]}{
         64(2s_2+1)\pi m_2}\,\,\kappa\,
  \sum_{\lambda^\prime}\,
  W_{\lambda^\prime,\lambda^\prime}(m_V,\omega)\,
  \rho^V_{\lambda^\prime,\lambda^\prime}(\theta_\ell)\,,
\label{eq:on-shell_correlated_polar_angle_distributions}\\
  \Gamma[X_2\to V X_1\to\ell^-\ell^+ X_1]
&=& \frac{{\rm Br}[V\to\ell^-\ell^+]}{
        16(2s_2+1)\pi m_2}\,\,\kappa\,
  \sum_{\lambda,\lambda_1}\,
  |\, {\cal C}_{\lambda,\lambda_1}(m_V)|^2\,,
\end{eqnarray}
with the constraint $|\lambda-\lambda_1|\leq s_2$ and the kinematical
factor $\kappa=\lambda^{1/2}(1, m^2_1/m^2_2, m^2_V/m^2_2)$.\s

The normalized invariant-mass and correlated polar-angle distributions,
which are valid for any value of the mass difference $m_2-m_1$ are
\begin{eqnarray}
         \frac{d N (m)}{d m}
\!\!  &=&\!\!
         \frac{\Pi_V(m)\,\, \kappa_*\,
         \sum_{\lambda,\lambda_1}|{\cal C}_{\lambda,\lambda_1}(m)|^2}{
         \int^{m_2-m_1}_{0} d m\, \Pi_V(m)\,\, \kappa_*
         \sum_{\lambda,\lambda_1}|{\cal C}_{\lambda,\lambda_1}(m)|^2}\,,
\label{eq:normalized_invariant_mass_distribution}\\[2mm]
\frac{d N}{d\cos\theta d\cos\theta_\ell}
\!\! &=&\!\! \frac{1}{4}\left[\, 1
                    + \frac{3}{2} A_\ell\, {\cal P}_V(m)\, \cos\omega
                                               \cos\theta_\ell
                    \right. \nonumber\\
&& + \left. \frac{1}{8} {\cal Q}_V(m)\, (3\cos^2\omega-1)\,
                                     (3\cos^2\theta_\ell -1)
                    \right]\,,
\label{eq:normalized_correlated_polar_angle_distribution}
\end{eqnarray}
where the $V$ propagator function $\Pi_V(m)$ and the $V$
longitudinal and tensor polarization components, ${\cal P}_V(m)$
and ${\cal Q}_V(m)$, are given by
\begin{eqnarray}
   \Pi_V(m)
&=& \frac{2 m^3}{(m^2-m^2_V)^2+m^2_V\Gamma^2_V}\,,
\label{eq:v_propagator_factor} \\
   {\cal P}_V(m)
&=& \frac{\sum_{\lambda_1}\left(|{\cal C}_{+,\lambda_1}|^2
                               -|{\cal C}_{-,\lambda_1}|^2\right)}{
          \sum_{\lambda,\lambda_1}|{\cal C}_{\lambda,\lambda_1}|^2}\,,
\label{eq:v_longitudinal_polarization}\\
   {\cal Q}_V(m)
&=& \frac{\sum_{\lambda_1}\left(|{\cal C}_{+,\lambda_1}|^2
                         -2|{\cal C}_{0,\lambda_1}|^2
                          +|{\cal C}_{-,\lambda_1}|^2)\right)}{
          \sum_{\lambda,\lambda_1}|{\cal C}_{\lambda,\lambda_1}|^2}\,,
\label{eq:v_tensor_polarization}
\end{eqnarray}
for a given value of the invariant mass $m$. \s

One crucial observation for the two-body decay $X_2\to V X_1$ involving
two Majorana particles, $X_2$ and $X_1$, is that the longitudinal
polarization ${\cal P}_V(m)$ is zero due to CPT invariance in the absence
of absorptive parts no matter of whether CP is broken or
not \cite{Choi:2003fs}. Therefore,
all the  normalized correlated polar-angle distributions are of a similar form
with a tensor polarization ${\cal Q}_V(m)$ encoding the information on the
spin and dynamical properties. Noting that there are many methods for
probing the general $X_2X_1V$ vertex, for example, through the pair
production of a non-diagonal $X_2X_1$ pair followed by the $X_2$ decay
into $X_1$ and SM leptons~\cite{Choi:2020spm}, for our specific numerical
demonstration in the present work, we investigate again a simple
two-body decay $X_2\to V X_1$ followed by a two-body decay
$V\to \ell^-\ell^+$ with $\ell=e$ or $\mu$, based on the couplings
listed in Table~\ref{tab:general_tensor_x2x1v_couplings}.\s

\subsection{Numerical investigations of the correlated distributions}
\label{seubsec:numerical_investigations_correlated_distributions}

Rather than performing a full-fledged analysis of the sequential decays for
every combination of the $X_2$ and $X_1$ spins, we restrict our present
numerical analysis to four spin combinations of $(s_2,s_1)=(0,0),\, (0,1),\, (1,0)$
and $(1,1)$ and set $V$ to be the gauge boson $Z$ with its SM couplings
to two leptons.\s

Specifically, for our numerical study, we consider two scenarios with the following sets of masses\,:
\begin{itemize}
\item \underline{Scenario 1} (${\cal S}$1)\,: $m_2=100\,{\rm GeV}$ and
      $m_1=30\,{\rm GeV}$ with an off-shell $V$.
\item \underline{Scenario 2} (${\cal S}$2)\,: $m_2=300\, {\rm GeV}$ and
      $m_1=100\, {\rm GeV}$ with an on-shell $V$.
\end{itemize}
with the $V$ mass and width set to the SM values of the $Z$ mass and width,
$m_V=m_Z=91.2\, {\rm GeV}$ and $\Gamma_V=\Gamma_Z= 2.5\, {\rm GeV}$~\cite{Zyla:2020zbs}.\s

For the couplings, we keep only the lowest-dimension terms in each spin combination
and assume the triple-vector coupling to be of a non-Abelian gauge group
type in the opposite normality and spin-[11] combination case.
As the normalized distributions are dependent only on the relative magnitudes
of couplings, we take in both scenarios
\begin{eqnarray}
b_1=c_1=d_1=d_2=\bar{a}_1=\bar{b}_1=\bar{c}_1=\bar{d}_1=1\quad
\mbox{and}\quad \bar{d}_2=-2\,,
\end{eqnarray}
while setting the other couplings to be zero, see
Table~\ref{tab:general_tensor_x2x1v_couplings}. We note that $\bar{d}_2=-2$ is
chosen for the triple-vector coupling to be of a trilinear coupling of gauge bosons.
In general the couplings themselves depend on the transferred momentum-squared
corresponding to the $V$ invariant mass-squared $m^2$. Nevertheless, we assume
them to be nearly constant as our focus is on the threshold behaviour quite close
to the invariant-mass endpoint of $m \simeq m_2-m_1$, which is $70\,{\rm GeV}$ in
our numerical example.\s

\vskip 0.5cm

\begin{figure}[H]
\begin{center}
\includegraphics[width=14.0cm, height=7.cm]{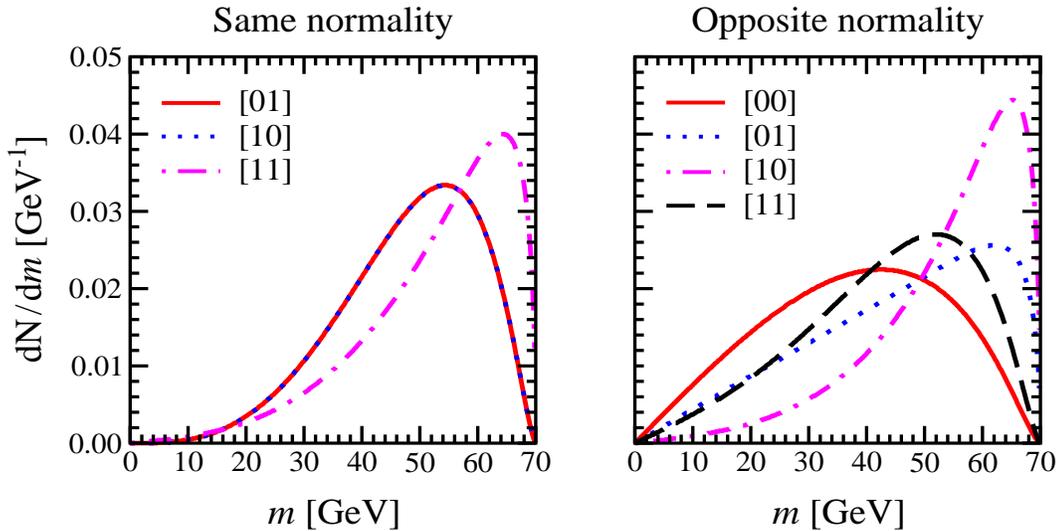}
\caption{\it The normalized invariant $V$ mass distribution in the same
normality case (Left) and in the opposite normality case (Right).
Depending on the spin values and normalities of two particles, $X_2$ and
$X_1$, the distribution shows its characteristic threshold behavior near
the invariant-mass end point of $m=m_2-m_1=70\,{\rm GeV}$ for
$m_2=100\, {\rm GeV}$ and $m_1=30\, {\rm GeV}$.
}
\label{fig:threshold_behavior_invariant_mass}
\end{center}
\end{figure}

Consistently with the threshold behaviors listed in the last column of
Table~\ref{tab:on_shell_tensor_polarization}, the same-normality $[11]$
and opposite-normality $[01]$ and $[10]$ invariant-mass spectra decreases
linearly with $\kappa_*\sim [(m_2-m_1)-m]^{1/2}$ and therefore steeply
just below the threshold, while the same-normality $[01]$ and $[10]$
and opposite-normality $[00]$ and $[11]$ invariant mass spectra decrease
in a cubic power of $\kappa_*$ with $\kappa^3_* \sim [(m_2-m_1)-m]^{3/2}$
and therefore rather gently as shown clearly in
Figure~\ref{fig:threshold_behavior_invariant_mass}.
Even with this distinct threshold pattern,
it is not possible to completely disentangle each spin-combination and
normality case, as the normalized same-normality $[01]$
and $[10]$ spectra are identical.\footnote{Numerically, we find that, if
$m_2$ is much larger than $m_{\rm max}=m_2-m_1$, the $[00]$ and $[11]$
distributions get indistinguishable, as the helicity-0 longitudinal mode of
the spin-1 $X_2$ contributes dominantly to the decay rate, consistently with
the equivalent Goldstone boson theorem~\cite{Cornwall:1974km,Vayonakis:1976vz,
Chanowitz:1985hj}.}
Therefore, it is necessary to utilize new independent observables for
a more clear disentanglement.\s

\vskip 0.5cm

\begin{figure}[H]
\begin{center}
\includegraphics[width=14.0cm, height=7.cm]{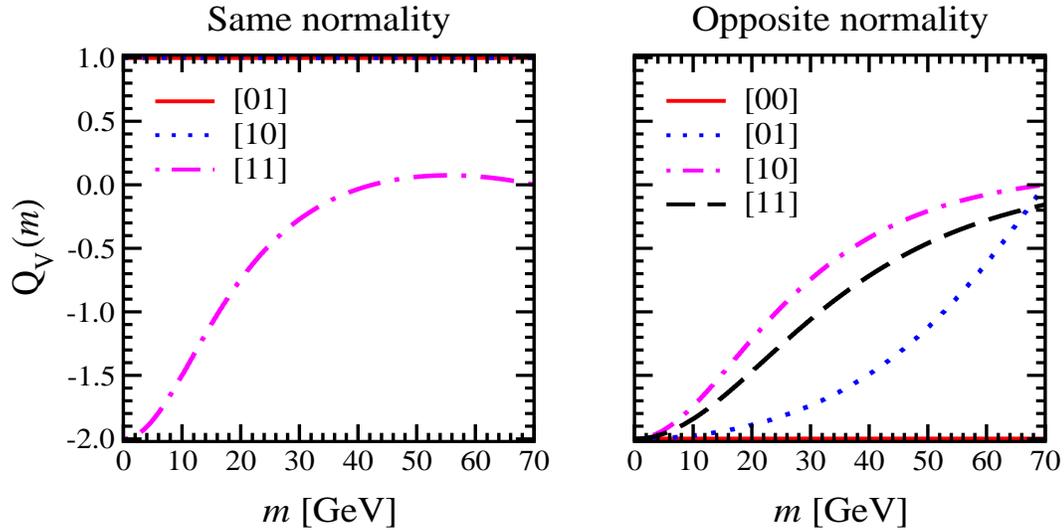}
\caption{\it The tensor polarization ${\cal Q}_V(m)$ as a function of the
invariant $V$ mass $m$ in the same normality case (Left) and
in the opposite normality case (Right). Depending on the spin values and
normalities of two particles, $X_2$ and $X_1$, the distribution shows
its characteristic $m$ dependence. In this numerical analysis,
$m_2=100\, {\rm GeV}$ and $m_1=30\,{\rm GeV}$ are taken.
}
\label{fig:qv_invariant_mass_distribution}
\end{center}
\end{figure}

In addition to the invariant-mass spectra, the tensor polarization
${\cal Q}_V(m)$ weighing the normalized correlated polar-angle distributions
as shown in Eq.$\,$(\ref{eq:normalized_correlated_polar_angle_distribution})
provides us with an additional handle for identifying the spin combination
and relative normalities. Figure~\ref{fig:qv_invariant_mass_distribution}
shows the dependence of the tensor polarization ${\cal Q}_V(m)$ on the
invariant mass $m$. In the same-normality case, the $[11]$ polar-angle
distribution can be clearly distinguished from the $[01]$ and $[10]$
polar-angle distributions, that are identical and constant, as a consequence
of the fact that the reduced helicity amplitudes ${\cal C}_{0,0}$
are vanishing. In the opposite normality case, all the four spin-combination
cases show different $m$-dependent behaviors. Specifically, ${\cal Q}_V(m)=-2$
in the $[00]$ case as only the longitudinal $V$ boson is produced.
Consequently, we find that, {\it although
not perfect, the invariant-mass threshold behaviors and correlated polar-angle
distributions enhance the resolution power for probing the spin and dynamical
properties of the particles $X_2$ and $X_1$.}\s

\vskip 0.5cm

\begin{table}[H]
\centering
\begin{tabular}{||c||c|c|c|c||c|c|c|c||} \hline
{\small Tensor Polarization}
& \multicolumn{4}{c||}{\small Same normality}
& \multicolumn{4}{c||}{\small Opposite normality}
  \\ \cline{2-9}
  {\color{blue} $[s_2\,\, s_1]$}
& {\color{blue} $[0\, 0]$}
& {\color{blue} $[0\, 1]$}
& {\color{blue} $[1\, 0]$}
& {\color{blue} $[1\, 1]$}
& {\color{blue} $[0\, 0]$}
& {\color{blue} $[0\, 1]$}
& {\color{blue} $[1\, 0]$}
& {\color{blue} $[1\, 1]$}
  \\ \hline
& & & & & & & & \\[-2mm]
  ${\cal Q}_V(m_V)$
& -- & $1.00$   & $1.00$  & $-0.32$
& $-2.00$  & $-1.66$  & $-0.70$ & $-1.08$
  \\[2mm]
  \hline
\end{tabular}
\vskip 0.3cm
\caption{\it The tensor polarization ${\cal Q}_V(m_V)$ for $m_2=300\,{\rm GeV}$,
$m_1=100\, {\rm GeV}$ and $m_V=m_Z=91.2\, {\rm GeV}$ for the combinations of the
$X_2$ and $X_1$ spins, $s_2$ and $s_1$, in the same normality case and in the
opposite normality case, respectively. }
\label{tab:on_shell_tensor_polarization}
\end{table}

If the mass difference $m_2-m_1$ is larger than the $V$ mass $m_V$, then
the vector boson $V$ is produced dominantly on-shell. In this situation,
the invariant-mass distribution is not available any more. Nevertheless,
the normalized correlated polar-angle distributions enable us to disentangle
the spin and normality combinations at least partially, as shown in
Table~\ref{tab:on_shell_tensor_polarization}.\s

As can be checked in Eq.$\,$(\ref{eq:normalized_correlated_polar_angle_distribution}),
the dependence of the correlated polar-angle correlations on the
polar-angle $\theta$ of the particle $V$ is encoded in the first and second Legendre polynomials, $P_1(\cos\omega)=\cos\omega$ and/or $P_2(\cos\omega)=(3\cos^2\omega-1)/2$, because the Wick helicity rotation angle $\omega$ is a function of the polar angle
$\theta$ as shown in Eqs.$\,$(\ref{eq:cos_wick_helicity_rotation_angle}) and
(\ref{eq:sin_wick_helicity_rotation_angle}). Furthermore, they depend on
the boost factor $\gamma_2=E_2/m_2$ denoting the energy $E_2$ of the decaying particle $X_2$ normalized to its mass $m_2$ in the LAB. As mentioned before,
the longitudinal polarization ${\cal P}_V(m)$ is zero due to CPT invariance
in the absence of absorptive parts. Therefore, the sensitivity of the
polar-angle distribution to each spin and normality scenario is determined
not only by the tensor polarization but also by the second Legendre
polynomial. \s

\vskip 0.5cm

\begin{figure}[H]
\begin{center}
\includegraphics[width=14.0cm, height=6.0cm]{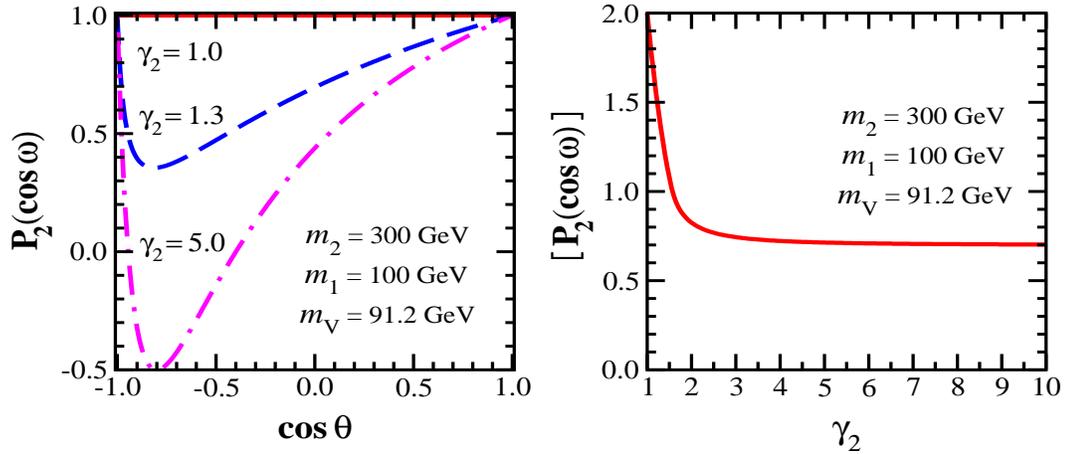}
\caption{\it (Left) The behavior of the second Legendre polynomial $P_2(\cos\omega)
=(3 \cos^2\omega-1)/2$ of the Wick helicity rotation angle $\omega$ as a
function of $\cos\theta$ for three values of the $X_2$ boost factor,
$\gamma_2=1, 1.3$ and $5$, corresponding to the $X_2$ speed, $\beta_2=0$,
$\beta_2< \beta$ and $\beta_2 > \beta$, respectively.
(Right) The $\gamma_2$ dependence of the integral $[P_2(\cos\omega)]
=\int^1_{-1} P_2(\cos\omega)\, d\cos\theta$. Numerically, $\beta\simeq 0.78$
for $m_2=300\, {\rm GeV}$,
$m_1=100\, {\rm GeV}$ and $m_V=m_Z=91.2\, {\rm GeV}$.
}
\label{fig:p2_cos_theta_integral_gamma_2}
\end{center}
\end{figure}

The left frame of Figure~\ref{fig:p2_cos_theta_integral_gamma_2} shows the
behavior of the second Legendre polynomial $P_2(\cos\omega)
=(3 \cos^2\omega-1)/2$ as an implicit function of $\cos\theta$ for three values
of the $X_2$ boost factor, $\gamma_2=1$ (red solid line), $1.3$ (blue
dashed line) and $5$ (magenta dot-dashed line), corresponding to
the $X_2$ speed, $\beta_2=0$, $\beta_2< \beta$ and $\beta_2 > \beta$,
respectively, and the right frame of Figure~\ref{fig:p2_cos_theta_integral_gamma_2}
shows the $\gamma_2$ dependence of the integral $[P_2(\cos\omega)]=\int^1_{-1} P_2(\cos\omega)\, d\cos\theta$.
Both of them are based on the scenario ${\cal S}$2 of
$m_2=300\, {\rm GeV}$, $m_1=100\, {\rm GeV}$ and
$m_V=m_Z=91.2\, {\rm GeV}$, in which $\beta\simeq 0.78$. The $\cos\theta$ distribution is greatly influenced by the value of $\gamma_2$,
when $\beta_2$ is less than $\beta$.
Actually, $\cos\omega=0$ at $\cos\theta=-\beta/\beta_2 \simeq -\beta\sim -0.78$
for $\gamma_2=5$, minimizing the second Legendre polynomial as shown by
the magenta dot-dashed line in the left frame. In contrast, the shape of
the curve changes so little for larger $\gamma_2$, as the value of
$\beta_2$ remains very close to unity. The single lepton polar-angle
distribution can be obtained by
integrating the second Legendre polynomial over the angle $\theta$, which is
still $\gamma_2$ dependent.
The right frame shows the monotonic decrease of the integral
$[P_2(\cos\omega)]$ converging asymptotically to a specific value.
Actually, the analytic expression of the asymptotic value for a given
$\beta$ is given by~\cite{Choi:2019aig}
\begin{eqnarray}
  {\cal L}_(\beta)
= \frac{1}{\beta^2}
  \left[\, 6-4\beta^2-\frac{3(1-\beta^2)}{\beta}
                     \ln\left(\frac{1+\beta}{1-\beta}\right)
  \,\right]\,,
\label{eq:asymptotic_value_integral_2nd_legendre_polynomial}
\end{eqnarray}
which is approximately $0.7$ for $\beta$ very close to 1 in the
scenario ${\cal S}2$ as shown in the right frame of
Figure~\ref{fig:p2_cos_theta_integral_gamma_2}. \s

{\it To summarize}, we have shown how the invariant-mass and/or correlated
polar-angle distributions can be expressed analytically in
a compact form by use of the Wick helicity rotation angle and polarization
functions and how they can be exploited efficiently for probing the $X_2X_1V$
vertex structure. Our restricted analysis is expected to be extended
straightforwardly to the much more general and sophisticated scenarios.\s

\section{Conclusions}
\label{sec:conclusions}

We have made a general and systematic study of the vector currents
of an on-shell or off-shell vector boson coupled to two integer-spin particles,
of which the masses and spins do not have to be identical.
The general vertex derived in a manifestly covariant formulation is
applicable independently of whether the particles are neutral or charged.
As a special case, we have probed in detail the case when
the two particles are Majorana bosons and then we have worked out explicitly
the constraints on the vertex due to discrete spacetime symmetries and
the Majorana condition valid for the Majorana bosons. \s

The general results obtained in a manifestly covariant form  have been
checked through the study of two-body decays $X_2\to V X_1$  of a heavier
Majorana boson $X_2$ into a lighter Majorana boson $X_1$ and an on-shell
or off-shell vector boson $V$ based on the helicity formalism
complementary to the covariant formalism  as demonstrated.\s

Considering two sequential 2-body decays, $X_2\to V X_1$
and $V \to \ell^-\ell^+$ with $\ell=e$ and $\mu$, we have investigated
how  the correlated polar-angle and/or invariant-mass distributions
enable us to determine the spin and dynamical structure of the triple
vertex fully. As a specific comparison, for all the combinations with
the spin value up to 1, we have found numerically that combining the
invariant-mass and polar-angle distributions allow us to characterize
the spin combinations effectively, although not perfect.\s

Although the half-integer spin case has to be worked out as well, this
general and model-independent study of the vector currents of two massive
(Majorana) particles of different masses and arbitrary integer spins
presented in the present work can be exploited for searching for new BSM
physics by probing various SM and BSM processes. Definitely, this work
can be expanded significantly for the general analysis of the triple
vertex of three particles of arbitrary spin.\footnote{The general triple
vertex of three particles of arbitrary integer spins has been described
and investigated in a different but powerful
formulation~\cite{Chung:1993da,Chung:1997jn,Chung:2003fk,Chung:2007nn}.
It will be valuable to compare and combine this formulation with
the manifestly covariant formulation adopted in the present work.}\s

\section*{Acknowledgment}
\label{sec:acknowledgment}

The work was in part by the Basic Science Research Program of Ministry of
Education through National Research Foundation of Korea
(Grant No. NRF-2016R1D1A3B01010529) and in part by the CERN-Korea theory
collaboration.


\begin{thebibliography}{99}


\bibitem{Glashow:1961tr}
S.~L.~Glashow,
``Partial Symmetries of Weak Interactions,''
Nucl. Phys. \textbf{22} (1961), 579-588
doi:10.1016/0029-5582(61)90469-2.

\bibitem{Weinberg:1967tq}
S.~Weinberg,
``A Model of Leptons,''
Phys. Rev. Lett. \textbf{19} (1967), 1264-1266
doi:10.1103/PhysRevLett.19.1264.

\bibitem{Salam:1968rm}
A.~Salam,
``Weak and Electromagnetic Interactions,''
Conf. Proc. C \textbf{680519} (1968), 367-377
doi:10.1142/9789812795915\_0034.

\bibitem{Majorana:1937vz}
E.~Majorana,
``Teoria simmetrica dell\textquoteright{}elettrone e del positrone,''
Nuovo Cim. \textbf{14} (1937), 171-184
doi:10.1007/BF02961314.

\bibitem{Dirac:1928hu}
P.~A.~M.~Dirac,
``The quantum theory of the electron,''
Proc. Roy. Soc. Lond. A \textbf{117} (1928), 610-624
doi:10.1098/rspa.1928.0023.


\bibitem{Aad:2012tfa}
G.~Aad \textit{et al.} [ATLAS],
``Observation of a new particle in the search for the Standard Model
Higgs boson with the ATLAS detector at the LHC,''
Phys. Lett. B \textbf{716} (2012), 1-29
doi:10.1016/j.physletb.2012.08.020
[arXiv:1207.7214 [hep-ex]].

\bibitem{Chatrchyan:2012ufa}
S.~Chatrchyan \textit{et al.} [CMS],
``Observation of a New Boson at a Mass of 125 GeV with the
CMS Experiment at the LHC,''
Phys. Lett. B \textbf{716} (2012), 30-61
doi:10.1016/j.physletb.2012.08.021
[arXiv:1207.7235 [hep-ex]].

\bibitem{Langacker:1980js}
P.~Langacker,
``Grand Unified Theories and Proton Decay,''
Phys. Rept. \textbf{72} (1981), 185
doi:10.1016/0370-1573(81)90059-4.

\bibitem{Croon:2019kpe}
D.~Croon, T.~E.~Gonzalo, L.~Graf, N.~Ko\v{s}nik and G.~White,
``GUT Physics in the era of the LHC,''
Front. in Phys. \textbf{7} (2019), 76
doi:10.3389/fphy.2019.00076
[arXiv:1903.04977 [hep-ph]].

\bibitem{Fayet:1976cr}
P.~Fayet and S.~Ferrara,
``Supersymmetry,''
Phys. Rept. \textbf{32} (1977), 249-334
doi:10.1016/0370-1573(77)90066-7.

\bibitem{Nilles:1983ge}
H.~P.~Nilles,
``Supersymmetry, Supergravity and Particle Physics,''
Phys. Rept. \textbf{110} (1984), 1-162
doi:10.1016/0370-1573(84)90008-5.

\bibitem{Haber:1984rc}
H.~E.~Haber and G.~L.~Kane,
``The Search for Supersymmetry: Probing Physics Beyond the Standard Model,''
Phys. Rept. \textbf{117} (1985), 75-263
doi:10.1016/0370-1573(85)90051-1.

\bibitem{GonzalezGarcia:2007ib}
M.~C.~Gonzalez-Garcia and M.~Maltoni,
``Phenomenology with Massive Neutrinos,''
Phys. Rept. \textbf{460} (2008), 1-129
doi:10.1016/j.physrep.2007.12.004
[arXiv:0704.1800 [hep-ph]].

\bibitem{Elliott:2014iha}
S.~R.~Elliott and M.~Franz,
``Colloquium: Majorana Fermions in nuclear, particle and solid-state physics,''
Rev. Mod. Phys. \textbf{87} (2015), 137
doi:10.1103/RevModPhys.87.137
[arXiv:1403.4976 [cond-mat.supr-con]].

\bibitem{Wilczek:2009np5614}
F.~Wilczek,
``Majorana returns,''
Nature Phys. \textbf{5} (2009), 614–618 (2009),
doi.org/10.1038/nphys1380.

\bibitem{Leggett:2016oci}
A.~J.~Leggett,
``Majorana fermions in condensed-matter physics,''
Int. J. Mod. Phys. B \textbf{30} (2016) no.19, 1630012
doi:10.1142/S0217979216300127.

\bibitem{Schechter:1981hw}
J.~Schechter and J.~W.~F.~Valle,
``Majorana Neutrinos and Magnetic Fields,''
Phys. Rev. D \textbf{24} (1981), 1883-1889
[erratum: Phys. Rev. D \textbf{25} (1982), 283]
doi:10.1103/PhysRevD.25.283.

\bibitem{Li:1981um}
L.~F.~Li and F.~Wilczek,
``Physical Processes Involving Majorana Neuntrinos,''
Phys. Rev. D \textbf{25} (1982), 143
doi:10.1103/PhysRevD.25.143.

\bibitem{Pal:1981rm}
P.~B.~Pal and L.~Wolfenstein,
``Radiative Decays of Massive Neutrinos,''
Phys. Rev. D \textbf{25} (1982), 766
doi:10.1103/PhysRevD.25.766.

\bibitem{Halprin:1983ez}
A.~Halprin, S.~T.~Petcov and S.~P.~Rosen,
``Effects of Light and Heavy Majorana Neutrinos in Neutrinoless Double Beta Decay,''
Phys. Lett. B \textbf{125} (1983), 335-338
doi:10.1016/0370-2693(83)91296-0.

\bibitem{Nieves:1982bq}
J.~F.~Nieves,
``Two Photon Decays of Heavy Neutrinos,''
Phys. Rev. D \textbf{28} (1983), 1664
doi:10.1103/PhysRevD.28.1664.

\bibitem{Khare:1983tm}
A.~Khare and J.~Oliensis,
``Constraints on the Interactions of Majorana Particles From \{CPT\} Invariance,''
Phys. Rev. D \textbf{29} (1984), 1542
doi:10.1103/PhysRevD.29.1542.

\bibitem{Bilenky:1984fg}
S.~M.~Bilenky, N.~P.~Nedelcheva and S.~T.~Petcov,
``Some Implications of the \{CP\} Invariance for Mixing of Majorana Neutrinos,''
Nucl. Phys. B \textbf{247} (1984), 61-69
doi:10.1016/0550-3213(84)90372-9.

\bibitem{Rosen:1983wu}
S.~P.~Rosen,
``General \{CP\} Properties of Neutrino Mass Eigenstates,''
[erratum: Phys. Rev. D \textbf{30} (1984), 1995]
doi:10.1103/PhysRevD.29.2535.

\bibitem{Kayser:1982br}
B.~Kayser,
``Majorana Neutrinos and their Electromagnetic Properties,''
Phys. Rev. D \textbf{26} (1982), 1662
doi:10.1103/PhysRevD.26.1662.

\bibitem{Kayser:1984ge}
B.~Kayser,
``CPT, CP, and C Phases and their Effects in Majorana Particle Processes,''
Phys. Rev. D \textbf{30} (1984), 1023
doi:10.1103/PhysRevD.30.1023.

\bibitem{Nieves:1996ff}
J.~F.~Nieves and P.~B.~Pal,
``Electromagnetic properties of neutral and charged spin 1 particles,''
Phys. Rev. D \textbf{55} (1997), 3118-3130
doi:10.1103/PhysRevD.55.3118
[arXiv:hep-ph/9611431 [hep-ph]].

\bibitem{Nieves:2013csa}
J.~F.~Nieves,
``Electromagnetic properties of spin-3/2 Majorana particles,''
Phys. Rev. D \textbf{88} (2013), 036006
doi:10.1103/PhysRevD.88.036006
[arXiv:1308.5889 [hep-ph]].

\bibitem{Radescu:1985wf}
E.~E.~Radescu,
``Comments on the Electromagnetic Properties of Majorana Fermions,''
Phys. Rev. D \textbf{32} (1985), 1266
doi:10.1103/PhysRevD.32.1266.

\bibitem{Boudjema:1988zs}
F.~Boudjema, C.~Hamzaoui, V.~Rahal and H.~C.~Ren,
``Electromagnetic Properties of Generalized Majorana Particles,''
Phys. Rev. Lett. \textbf{62} (1989), 852
doi:10.1103/PhysRevLett.62.852.

\bibitem{Boudjema:1990st}
F.~Boudjema and C.~Hamzaoui,
``Massive and massless Majorana particles of arbitrary spin: Covariant gauge couplings and production properties,''
Phys. Rev. D \textbf{43} (1991), 3748-3758
doi:10.1103/PhysRevD.43.3748.

\bibitem{Ellis:1983er}
J.~R.~Ellis, J.~M.~Frere, J.~S.~Hagelin, G.~L.~Kane and S.~T.~Petcov,
``Search for Neutral Gauge Fermions in $e^+ e^-$ Annihilation,''
Phys. Lett. B \textbf{132} (1983), 436-442
doi:10.1016/0370-2693(83)90343-X.

\bibitem{Petcov:1984nf}
S.~T.~Petcov,
``Possible Signature for Production of Majorana Particles in $e^+ e^-$ and $p \bar{p}$ Collisions,''
Phys. Lett. B \textbf{139} (1984), 421-426
doi:10.1016/0370-2693(84)91844-6.

\bibitem{Bilenky:1985wu}
S.~M.~Bilenky, N.~P.~Nedelcheva and E.~K.~Khristova,
``On Production of Majorana Particles in Polarized $e^+ e^-$ Collisions,''
Phys. Lett. B \textbf{161} (1985), 397-399
doi:10.1016/0370-2693(85)90786-5.

\bibitem{Bilenky:1986nd}
S.~M.~Bilenky, E.~K.~Khristova and N.~P.~Nedelcheva,
``Possible Tests for Majorana Nature of Heavy Neutral Fermions Produced
in Polarized $e^+ e^-$ Collisions,''
Bulg. J. Phys. \textbf{13} (1986), 283
JINR-E2-86-353.

\bibitem{Petcov:1986sc}
S.~T.~Petcov,
``\{CP\} Violation Effect in Neutralino Pair Production in $e^+ e^-$ Annihilation and the Electric Dipole Moment of the Electron,''
Phys. Lett. B \textbf{178} (1986), 57-64
doi:10.1016/0370-2693(86)90469-7.

\bibitem{MoortgatPick:2002iq}
G.~A.~Moortgat-Pick and H.~Fraas,
``Influence of CP and CPT on production and decay of Dirac and Majorana fermions,''
Eur. Phys. J. C \textbf{25} (2002), 189-197
doi:10.1007/s10052-002-0979-x
[arXiv:hep-ph/0204333 [hep-ph]].

\bibitem{Khristova:1987xq}
E.~K.~Khristova and N.~P.~Nedelcheva,
``On the Lightest Supersymmetric Particle in Polarized $e^+ e^-$ Collisions,''
Phys. Lett. B \textbf{208} (1988), 525-529
doi:10.1016/0370-2693(88)90661-2.

\bibitem{Balantekin:2018ukw}
A.~B.~Balantekin, A.~de Gouv\^ea and B.~Kayser,
``Addressing the Majorana vs. Dirac Question with Neutrino Decays,''
Phys. Lett. B \textbf{789} (2019), 488-495
doi:10.1016/j.physletb.2018.11.068
[arXiv:1808.10518 [hep-ph]].

\bibitem{Renard:1981es}
F.~M.~Renard,
``Tests of Neutral Gauge Boson Self Couplings with $e^+ e^- \to \gamma Z$,''
Nucl. Phys. B \textbf{196} (1982), 93-108
doi:10.1016/0550-3213(82)90304-2.

\bibitem{Hagiwara:1986vm}
K.~Hagiwara, R.~D.~Peccei, D.~Zeppenfeld and K.~Hikasa,
``Probing the Weak Boson Sector in $e^+ e^- \to W^+ W^-$,''
Nucl. Phys. B \textbf{282} (1987), 253-307
doi:10.1016/0550-3213(87)90685-7.

\bibitem{Choudhury:1994nt}
D.~Choudhury and S.~D.~Rindani,
``Test of CP violating neutral gauge boson vertices in $e^+ e^- \to \gamma Z$,''
Phys. Lett. B \textbf{335} (1994), 198-204
doi:10.1016/0370-2693(94)91413-3
[arXiv:hep-ph/9405242 [hep-ph]].

\bibitem{Ananthanarayan:2004eb}
B.~Ananthanarayan, S.~D.~Rindani, R.~K.~Singh and A.~Bartl,
``Transverse beam polarization and CP-violating triple-gauge-boson couplings in $e^+ e^-\to \gamma Z$,''
Phys. Lett. B \textbf{593} (2004), 95-104
[erratum: Phys. Lett. B \textbf{608} (2005), 274-275]
doi:10.1016/j.physletb.2005.01.009
[arXiv:hep-ph/0404106 [hep-ph]].

\bibitem{Rahaman:2016pqj}
R.~Rahaman and R.~K.~Singh,
``On polarization parameters of spin-1 particles and anomalous couplings in $e^+e^-\rightarrow ZZ/Z\gamma $,''
Eur. Phys. J. C \textbf{76} (2016) no.10, 539
doi:10.1140/epjc/s10052-016-4374-4
[arXiv:1604.06677 [hep-ph]].

\bibitem{Rahaman:2017qql}
R.~Rahaman and R.~K.~Singh,
``On the choice of beam polarization in $e^+e^-\rightarrow ZZ/Z\gamma $ and anomalous triple gauge-boson couplings,''
Eur. Phys. J. C \textbf{77} (2017) no.8, 521
doi:10.1140/epjc/s10052-017-5093-1
[arXiv:1703.06437 [hep-ph]].

\bibitem{Rahaman:2020jll}
R.~Rahaman,
``Study of anomalous gauge boson self-couplings and the role of spin-$1$ polarizations,''
[arXiv:2007.07649 [hep-ph]].

\bibitem{Rahaman:2018ujg}
R.~Rahaman and R.~K.~Singh,
``Anomalous triple gauge boson couplings in $ZZ$ production at the LHC and the role of $Z$ boson polarizations,''
Nucl. Phys. B \textbf{948} (2019), 114754
doi:10.1016/j.nuclphysb.2019.114754
[arXiv:1810.11657 [hep-ph]].

\bibitem{Choi:2015zka}
S.~Y.~Choi, T.~Han, J.~Kalinowski, K.~Rolbiecki and X.~Wang,
``Characterizing invisible electroweak particles through single-photon processes at high energy $e^+e^-$ colliders,''
Phys. Rev. D \textbf{92} (2015) no.9, 095006
doi:10.1103/PhysRevD.92.095006
[arXiv:1503.08538 [hep-ph]].

\bibitem{Choi:2001ww}
S.~Y.~Choi, J.~Kalinowski, G.~A.~Moortgat-Pick and P.~M.~Zerwas,
``Analysis of the neutralino system in supersymmetric theories,''
Eur. Phys. J. C \textbf{22} (2001), 563-579
doi:10.1007/s100520100808
[arXiv:hep-ph/0108117 [hep-ph]].

\bibitem{Choi:1994nv}
S.~Y.~Choi,
``Probing the weak boson sector in $\gamma e \to Z e$,''
Z. Phys. C \textbf{68} (1995), 163-172
doi:10.1007/BF01579815
[arXiv:hep-ph/9412300 [hep-ph]].

\bibitem{Choi:2003fs}
S.~Y.~Choi and Y.~G.~Kim,
``Analysis of the neutralino system in two body decays of neutralinos,''
Phys. Rev. D \textbf{69} (2004), 015011
doi:10.1103/PhysRevD.69.015011
[arXiv:hep-ph/0311037 [hep-ph]].

\bibitem{Choi:2005gt}
S.~Y.~Choi, B.~C.~Chung, J.~Kalinowski, Y.~G.~Kim and K.~Rolbiecki,
``Analysis of the neutralino system in three-body leptonic decays of neutralinos,''
Eur. Phys. J. C \textbf{46} (2006), 511-520
doi:10.1140/epjc/s2006-02482-1
[arXiv:hep-ph/0504122 [hep-ph]].

\bibitem{Choi:2018sqc}
S.~Y.~Choi,
``$Z$-boson polarization as a model-discrimination analyzer,''
Phys. Rev. D \textbf{98} (2018) no.11, 115037
doi:10.1103/PhysRevD.98.115037
[arXiv:1811.10377 [hep-ph]].

\bibitem{Choi:2019aig}
S.~Y.~Choi, J.~H.~Jeong and J.~H.~Song,
``General Spin Analysis from Angular Correlations in Two-Body Decays,''
Eur. Phys. J. Plus \textbf{135} (2020) no.2, 210
doi:10.1140/epjp/s13360-020-00132-1
[arXiv:1903.00166 [hep-ph]].

\bibitem{Choi:2020spm}
S.~Y.~Choi and J.~H.~Jeong,
``A Nondiagonal Pair of Majorana Particles at $e^+e^-$ Colliders,''
[arXiv:2012.14613 [hep-ph]].

\bibitem{Jacob:1959at}
M.~Jacob and G.~C.~Wick,
``On the General Theory of Collisions for Particles with Spin,''
Annals Phys. \textbf{7} (1959), 404-428
doi:10.1016/0003-4916(59)90051-X.

\bibitem{Wick:1962zz}
G.~C.~Wick,
``Angular momentum states for three relativistic particles,''
Annals Phys. \textbf{18} (1962), 65-80
doi:10.1016/0003-4916(62)90059-3.

\bibitem{Leader:2001gr}
E.~Leader,
``Spin in particle physics,''
Camb. Monogr. Part. Phys. Nucl. Phys. Cosmol. \textbf{15} (2011),
pp.1-500.

\bibitem{Chung:1971ri}
S.~U.~Chung,
``SPIN FORMALISMS,''
doi:10.5170/CERN-1971-008.

\bibitem{Behrends:1957rup}
R.~E.~Behrends and C.~Fronsdal,
``Fermi Decay of Higher Spin Particles,''
Phys. Rev. \textbf{106} (1957) no.2, 345
doi:10.1103/PhysRev.106.345.

\bibitem{Auvil:1966eao}
P.~R.~Auvil and J.~J.~Brehm,
``Wave Functions for Particles of Higher Spin,''
Phys. Rev. \textbf{145} (1966) no.4, 1152
doi:10.1103/PhysRev.145.1152.

\bibitem{Weinberg:1995mt}
S.~Weinberg,
``The Quantum theory of fields. Vol. 1: Foundations,''
(Cambridge University Press, 1995), ISBN 0-521-55001-7.

\bibitem{Landau:1948kw}
L.~D.~Landau,
``On the angular momentum of a system of two photons,''
Dokl. Akad. Nauk SSSR \textbf{60} (1948) no.2, 207-209
doi:10.1016/B978-0-08-010586-4.50070-5.

\bibitem{Yang:1950rg}
C.~N.~Yang,
``Selection Rules for the Dematerialization of a Particle Into Two Photons,''
Phys. Rev. \textbf{77} (1950), 242-245
doi:10.1103/PhysRev.77.242.

\bibitem{merose2011}
M.~E.~Rose,
``Elementary Theory of Angular Momentum" (Dover Publication Inc., New York, 2011)
ISBN-13: 978-0486684802.

\bibitem{Zyla:2020zbs}
P.~A.~Zyla \textit{et al.} [Particle Data Group],
``Review of Particle Physics,''
PTEP \textbf{2020} (2020) no.8, 083C01
doi:10.1093/ptep/ptaa104.

\bibitem{Cornwall:1974km}
J.~M.~Cornwall, D.~N.~Levin and G.~Tiktopoulos,
``Derivation of Gauge Invariance from High-Energy Unitarity Bounds on the s Matrix,''
Phys. Rev. D \textbf{10} (1974), 1145
[erratum: Phys. Rev. D \textbf{11} (1975), 972]
doi:10.1103/PhysRevD.10.1145.

\bibitem{Vayonakis:1976vz}
C.~E.~Vayonakis,
``Born Helicity Amplitudes and Cross-Sections in Nonabelian Gauge Theories,''
Lett. Nuovo Cim. \textbf{17} (1976), 383
doi:10.1007/BF02746538.

\bibitem{Chanowitz:1985hj}
M.~S.~Chanowitz and M.~K.~Gaillard,
``The TeV Physics of Strongly Interacting W's and Z's,''
Nucl. Phys. B \textbf{261} (1985), 379-431
doi:10.1016/0550-3213(85)90580-2.

\bibitem{Chung:1993da}
S.~U.~Chung,
``Helicity coupling amplitudes in tensor formalism,''
Phys. Rev. D \textbf{48} (1993), 1225-1239
[erratum: Phys. Rev. D \textbf{56} (1997), 4419]
doi:10.1103/PhysRevD.56.4419.

\bibitem{Chung:1997jn}
S.~U.~Chung,
``A General formulation of covariant helicity coupling amplitudes,''
Phys. Rev. D \textbf{57} (1998), 431-442
doi:10.1103/PhysRevD.57.431.

\bibitem{Chung:2003fk}
S.~U.~Chung,
``Covariant helicity-coupling amplitudes: Principles and examples,''
Int. J. Mod. Phys. A \textbf{18} (2003), 457-473
doi:10.1142/S0217751X03014381.

\bibitem{Chung:2007nn}
S.~U.~Chung and J.~Friedrich,
``Covariant helicity-coupling amplitudes: A New formulation,''
Phys. Rev. D \textbf{78} (2008), 074027
doi:10.1103/PhysRevD.78.074027
[arXiv:0711.3143 [hep-ph]].


\end{thebibliography}
\end{document}